\documentclass[11pt]{article}

\usepackage[margin=1in]{geometry}
\usepackage{amsmath,amssymb,amsthm,mathtools}
\usepackage{braket}
\usepackage{enumitem}
\usepackage{graphicx}
\usepackage{authblk}
\usepackage{tikz-cd}
\usepackage[hidelinks]{hyperref}
\usepackage[maxnames=3,backend=biber,style=numeric-comp,sorting=none,url=false]{biblatex}
\graphicspath{{Figures/}}

\newcommand{\R}{\mathbb{R}}

\newcommand{\Irr}{\operatorname{Irr}}

\newcommand{\proseparagraphs}{%
\setlength{\parindent}{15pt}%
\setlength{\parskip}{0pt}%
}
\newcommand{\mathematicalparagraphs}{%
\setlength{\parindent}{0pt}%
\setlength{\parskip}{\baselineskip}%
}

\newtheorem{definition}{Definition}[section]
\newtheorem{proposition}[definition]{Proposition}
\newtheorem{theorem}[definition]{Theorem}

\newtheorem{example}[definition]{Example}

\title{A Geometric Framework for Local-to-Global Symmetry: Global Constraints on Local Dynamics}
\author[1]{Connor Terry Weatherly\thanks{Email: \href{mailto:connorweatherly@gmail.com}{connorweatherly@gmail.com}}}
\author[1]{Roel Tempelaar\thanks{Corresponding author: \href{mailto:roel.tempelaar@northwestern.edu}{roel.tempelaar@northwestern.edu}}}
\affil[1]{Department of Chemistry, Northwestern University, Evanston, Illinois 60208-3113, USA}
\date{}

\begin{document}
\mathematicalparagraphs
\maketitle

\begin{abstract}
  Aggregate molecular systems are routinely understood by decomposing them into local structural units, such as monomers in J/H-aggregates, chemical moieties, motifs, and functional groups. Implicit in such decompositions is the assumption that these local structures remain meaningful within the aggregate. Here, we ask how the symmetries of constituent structures relate to aggregate symmetry, and how this relation determines the constraints that global symmetry imposes on transitions between local symmetry sectors. We construct a covariant local-to-global representation from framed local symmetry data and show that its reconstruction map intertwines with a global quantum representation. By identifying which global symmetry sectors overlap each local symmetry sector, we resolve symmetry-enforced dynamical exclusions into site-resolved pathways. Through Frobenius reciprocity, these exclusions can be determined from standard character tables of the local and global symmetry groups, with the character values in the inner product paired according to the relative positions and orientations of local structures within the aggregate. Although motivated by molecular aggregates, the construction is formulated for arbitrary geometries modeled by a homogeneous space of an ambient transformation group.
\end{abstract}

\section{Introduction}
\proseparagraphs
Understanding wholes through their parts is a central methodological principle of science. Its prominence within chemistry is immediate, where matter is decomposed into atoms, molecules, and larger complexes whose physicochemical properties are classified.\cite{hendry23_struc_essen_exist_chemis,talanquer18_progr_reason_about_struc_proper_relat} Combined with physical theory, this taxonomy provides a reliable framework for predicting the structural, optical, electrochemical, and reactive properties of assemblies from their molecular components.\cite{katritzky10_quant_correl_physic_chemic_proper,musil21_physic_inspir_struc_repres_molec_mater}

Yet this reductionist picture has an inherent contingency: physicochemical properties attributed to molecular components are seldom wholly independent of their environment.\cite{primas83_chemis_quant_mechan_reduc,hendry23_struc_essen_exist_chemis} Chemical inference nevertheless succeeds because distinct molecules are often effectively separable, owing to rapidly decaying spatial and temporal correlations.\cite{nitzan2013chemical,prodan05_nears_elect_matter,fias17_chemic_trans_funct_group_follow} Molecular aggregates challenge this separability, in that strong electronic connectivity establishes structural and dynamical correlations between their constituent moieties.\cite{hestand18_expan_theor_h_j_molec_aggreg,KUHN201147} Despite the chemist's inclination to decompose such systems into simple subunits, that the properties of isolated subunits remain meaningful within the aggregate is, as such, not always obvious.\cite{primas83_chemis_quant_mechan_reduc}

Among the physicochemical properties used to classify molecules, symmetry occupies a unique role. Rather than assigning a numerical attribute, it organizes states and operators representation-theoretically. Systems carrying equivalent symmetry actions therefore share symmetry-constrained statements---including criteria for allowed and forbidden transitions---independently of their microscopic realization.\cite{wigner2013group,eckart30_applic_group_theor_to_quant,Berndt2007,woit17_quant_theor_group_repres}

Chemical intuition concerning molecular complexes frequently begins with the observation of symmetries within local structural moieties. It is the meaning of such local symmetries within the aggregate that we deliberate herein. In particular, we ask under what conditions local symmetry information survives the passage to the aggregate in a manner that permits sharper symmetry-based inferences than those obtainable from the aggregate symmetry alone.

Answering this question requires a mathematical framework relating local and global descriptions of molecular systems.

Full aggregate electronic-structure calculations apply quantum theory to
predict properties of particular molecular configurations.\cite{garzon-ramirez26_how_symmet_gover_dihed_angle,peeks17_elect_deloc_radic_cation_porph,plasser12_analy_excit_charg_trans_inter} Such predictions
do not by themselves establish how properties of constituent
subunits and their interactions determine behavior across different aggregates.
Synthesizing such recurring properties into general principles requires abstracting components and interactions into models that organize aggregate behavior and support inferences across systems.

Site-based Hamiltonians provide such a component-based approach. Frenkel-exciton, Holstein, and related models decompose an aggregate into interacting sites, making localization, transport, coherence, and relations between subunits naturally expressible.\cite{hestand18_expan_theor_h_j_molec_aggreg} Such descriptions underpin models of molecular aggregates, molecular electronics, quantum materials, and spin-dependent phenomena.\cite{KUHN201147,nitzan03_elect_trans_molec_wire_junct,dagotto94_correl_elect_high_temper_super,guo12_spin_selec_trans_elect_dna_doubl_helix} Structural symmetry has also been exploited to constrain and simplify aggregate dynamics. For example, permutation symmetry in identically coupled molecular aggregates has been used to reduce the complexity of many-body dynamics through a bosonic representation of the permutation-symmetric sector.\cite{raghavan-chitra26_permut_symmet_molec_aggreg} Other approaches, notably those developed by Spano and collaborators,
exploit translational and crystal symmetries in specific aggregate models
to construct symmetry-adapted exciton–vibrational states from molecular
electronic and vibrational states of specified symmetry, and thereby
determine polarization-dependent optical selection rules.\cite{spano03_fundam_photop_conjug_oligom_herrin_aggreg,hestand18_expan_theor_h_j_molec_aggreg,spano10_vibron_coupl_j_aggreg_beyon,spano91_excit_confin_geomet}

These developments demonstrate the importance of symmetry and local geometry in aggregate models, but do not establish a general representation-theoretic relation between local and global symmetries. Such a relation is particularly relevant to problems such as chirality-induced spin selectivity, where local geometric orientation must be related to spin dynamics consistently with the geometry of the complete molecular system.\cite{evers22_theor_chiral_induc_spin_selec} A decomposition alone identifies the relevant subobjects, but does not itself provide a covariant relation among their local frames, local symmetry sectors, global transformations, and site permutations.

\begin{figure}[htbp]
\centering
\includegraphics[width=0.5\linewidth]{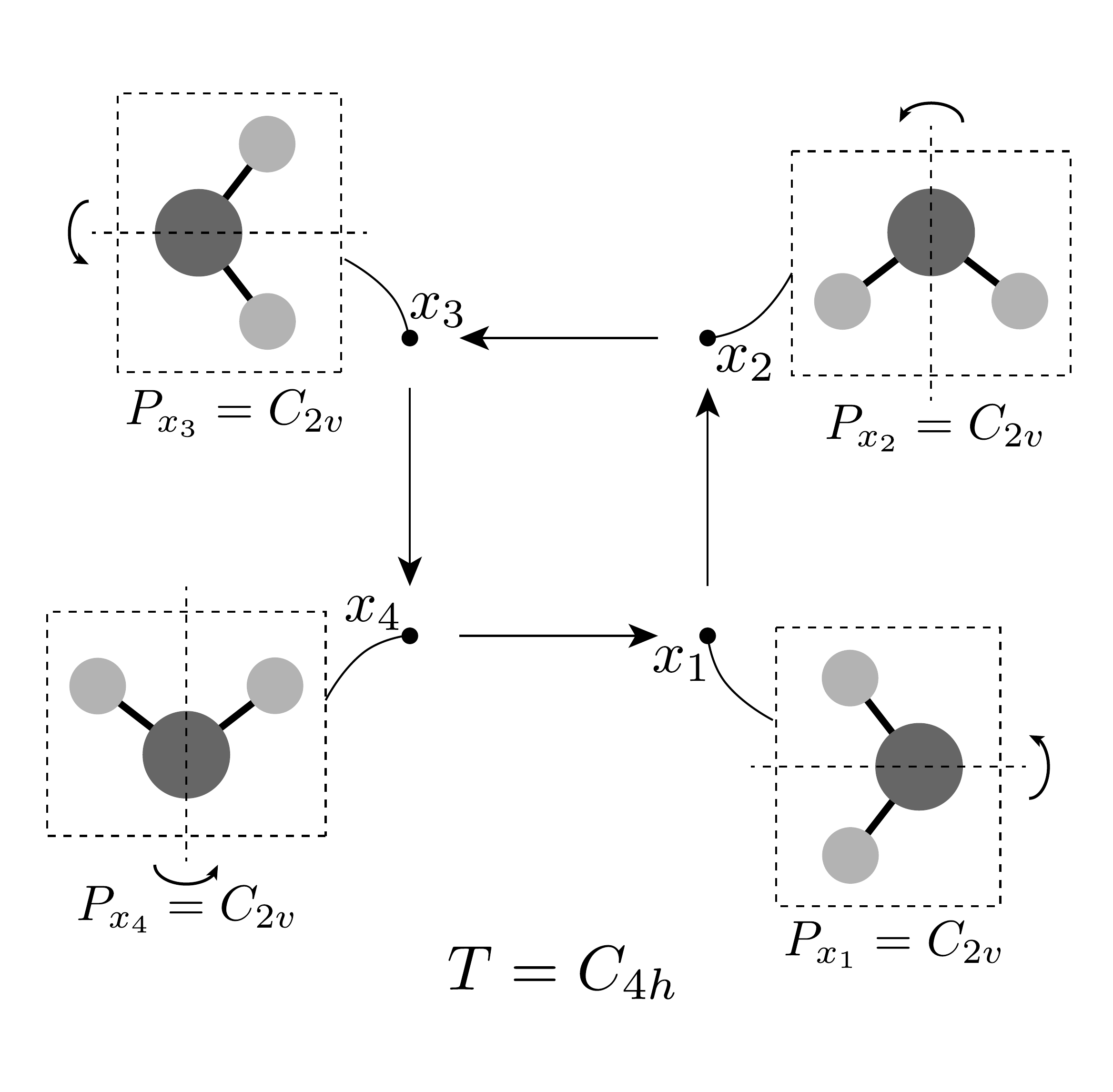}
\caption{Framed local structural symmetry groups on a four-site configuration. Although each site carries a local structural symmetry group $P_{x_i} = C_{2v}$, its symmetry elements are defined relative to a distinct local frame and are compared through transport between sites.}
\label{fig:framed-point-groups}
\end{figure}

To address this, we develop a covariant representation-theoretic framework relating local and global symmetry through geometric structure. Although motivated by molecular aggregates, the construction is formulated for arbitrary geometries modeled by a homogeneous space of an ambient transformation group $G$. The framework uses established constructions from principal-bundle theory and representation theory. The central mathematical result of this work is the determination of the maximal subgroup of $G$ that preserves framed local-symmetry data, such as the symmetries of structural subunits of a molecular aggregate (see Figure \ref{fig:framed-point-groups}). We provide a practical definition of this maximal symmetry group so that it may be calculated from a finite set of local symmetry data (such as the point groups and orientations of molecular subunits). We then demonstrate how representations of each framed local structural symmetry group assemble into a global representation--independent up to equivalence of the choice of fiber-coordinate identifications--and characterize the relation between its local and global symmetry sectors. As a principal application, this structure resolves the globally invariant blocks of a Hamiltonian into site-to-site symmetry channels. Character scalar products and Frobenius reciprocity\cite{serre77_linear_repres_finit_group} identify excluded local pathways, while the channel norms quantify the couplings in a specified Hamiltonian.

A particularly relevant antecedent is the site-symmetry approach of Evarestov and Smirnov, which relates site-centered and globally symmetry-adapted orbitals through representations induced from stabilizers within a prescribed molecular or crystal symmetry group.\cite{evarestov84_applic_band_repres_space_group,evarestov97_site_symmet_cryst} Topological quantum chemistry develops this approach further, connecting localized orbital symmetry to band representations, momentum-space connectivity, and obstructions to symmetry-compatible Wannier descriptions.\cite{bradlyn17_topol_quant_chemis,cano18_build_block_topol_quant_chemis,cano21_band_repres_topol_quant_chemis} These constructions begin with the global symmetry group and representations of its site stabilizers. Here, we instead begin with framed local structural data, determine the compatible global group, and assemble local model representations covariantly, with physical reconstruction provided separately.

The manuscript is organized as follows. Section~\ref{sec:semantics} introduces the geometric
and quantum ingredients and their physical interpretation. Section~\ref{sec:geometry}
constructs the geometric connection between local structural symmetries and
the compatible global symmetry group. Section~\ref{sec:representation} develops the associated
local-to-global representation and relates it to the global quantum
description. Section~\ref{sec:sector-compatibility} establishes the compatibility of local and global
symmetry sectors and its consequences for invariant operators. Section~\ref{sec:application}
applies the framework to Hamiltonian symmetrization, coupling constraints,
and dynamics in a square aggregate. Sections~\ref{sec:discussion} and \ref{sec:conclusion} discuss the
implications and summarize the results.

\section{Chemical and Physical Semantics}
\label{sec:semantics}
\proseparagraphs
This section introduces the intended physical interpretation of the geometric and quantum structures used throughout the paper. The formal constructions that follow are stated separately from this interpretation. Our aim is for the theory to proceed by deduction from explicitly stated mathematical assumptions encoding physical premises, such as the symmetry of an initial state. The resulting conclusions are then interpreted through the correspondence between mathematical syntax and physical semantics.

\subsection{Geometry}
\begin{figure}[!htb]
\centering
\includegraphics[width=0.4\linewidth]{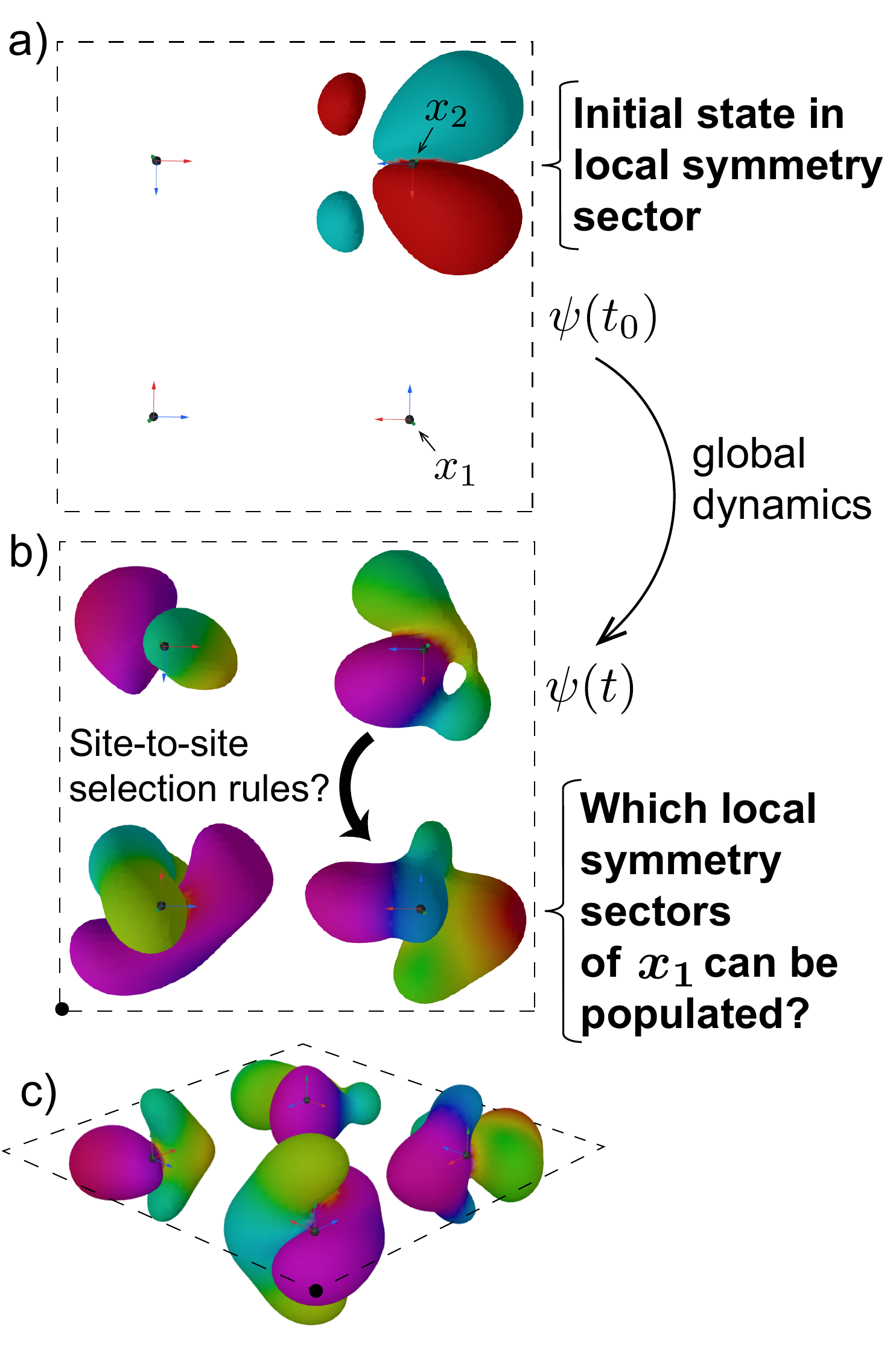}
\caption{Local symmetry and global dynamics in the four-site aggregate shown in Figure~\ref{fig:framed-point-groups}. (a) The initial state $\psi(t_0)$ is localized at site $x_2$ in the local symmetry sector labeled by the irrep symbol $B_1$. (b) Evolution under the aggregate Hamiltonian produces a delocalized state $\psi(t)$. The figure asks whether symmetry enforces site-to-site selection rules, such as requiring the population of a local symmetry sector at site \(x_1\) to remain zero throughout the evolution. (c) A rotated view of the same evolved state shown in (b).}
\label{fig:square-orbitals}
\end{figure}
We consider an aggregate as a collection of spatially situated geometric objects constituting the subunits of a whole geometric object. For molecular systems these subunits are chosen structural moieties--groupings of the total nuclear geometry into components. Each subunit possesses a local structural symmetry group composed of transformations about a fixed point, which we call a \emph{site}, such as indicated by $x_1, x_2$, etc. in Figure~\ref{fig:framed-point-groups}. In Figure~\ref{fig:framed-point-groups} each site represents a local molecular environment together with a chosen local frame. The corresponding local structural symmetry group \(P_x\) describes the symmetries of the site structure in isolation from the aggregate, relative to its own frame. For example, the four identical water like structures in Figure~\ref{fig:framed-point-groups} each have a local structural symmetry group $P_{x} = C_{2v}$. Importantly, we distinguish the local structural symmetry group $P_x$ from the aggregate site stabilizer $T_x=\{g\in T\mid g\cdot x=x\}$, conventionally called the site-symmetry group in crystallographic treatments. Geometric objects are specified by a labeling of each site. The global symmetry group \(T\) acts by permuting sites sharing the same label, while simultaneously transporting their local frames. Consequently, local symmetry data cannot be compared directly between different sites, but must first be related through the geometric transport determined by the framed geometry of the sites.

\subsection{Quantum Representations}
If a subunit of an aggregate is taken in isolation, its quantum state space carries a unitary representation of its symmetry group, with which the Hamiltonian commutes. For a compact symmetry group, the Hilbert space decomposes into isotypic components (symmetry sectors) labeled by irreducible representations.\cite{Berndt2007,serre77_linear_repres_finit_group} These components are invariant under the dynamics generated by the Hamiltonian, and when the Hamiltonian has a complete set of eigenvectors, it admits an eigenbasis adapted to this isotypic decomposition.\cite{woit17_quant_theor_group_repres}

If a quantum state within an aggregate is sufficiently localized to a subunit, such that its local potential can be approximated by the subunit structure, then its local symmetry classification may remain meaningful. The dynamics, however, are constrained by the symmetry of the complete aggregate and need not preserve the local symmetry classification. A state initially localized on a particular subunit, characterized by an irrep of its symmetry group, may evolve beyond that local symmetry sector and delocalize across other subunits. This is illustrated in Figure~\ref{fig:square-orbitals}. Figure~\ref{fig:square-orbitals}a shows a state initially localized on site \(x_2\), characterized by the \(B_1\) irrep of its local structural symmetry group. Figure~\ref{fig:square-orbitals}b shows the state following its evolution under a Hamiltonian invariant under the global \(C_{4h}\) symmetry group, resulting in its delocalization across the aggregate. Figure~\ref{fig:square-orbitals}c presents the same state from a rotated perspective. In the delocalized state, the symmetry classification relative to the individual subunits is no longer visually apparent. Nevertheless, the local structural symmetries remain, raising the central question whether these local symmetries continue to impose constraints on the dynamics of the aggregate? More precisely, can a state initially characterized by one local irrep evolve to populate a specified symmetry sector at another site, or are some such transitions forbidden by symmetry? This question motivates the local-to-global representation framework developed herein.

\section{Geometric Connection Between Local and Global Symmetry}
\label{sec:geometry}
\mathematicalparagraphs
We now formalize the geometric data introduced above. The construction
specifies the site positions and a local frame at each site, allowing local
structural symmetries to be compared within a common geometry. We then
determine which ambient transformations preserve this structure and hence
constitute compatible global symmetries. Homogeneous spaces describe the
possible site positions, while representatives in the ambient transformation
group specify the local frames.

\subsection{Sites and local structural symmetry groups}
\label{sec:sites-local-groups}
Let \(G\) be an ambient transformation group and \(H\leq G\) a reference
stabilizer subgroup, specifying the ambient local transformations at a
reference site. The homogeneous space \(G/H\) describes the possible site
positions,\cite[p.~4]{Berndt2007} with reference point \(x_0=eH\) satisfying
\[
H=\operatorname{Stab}_G(x_0)
:=\{g\in G\mid g\cdot x_0=x_0\},
\]
under the action \(g\cdot(aH)=(ga)H\). Introduce, now, a finite set
\(X\subset G/H\) specifying the site positions. Associated with each
\(x\in X\) is a local structural symmetry group \(P_x\leq H\).

\begin{example}
\label{exp:E3-quotient-structure}
  Consider, for example, the Euclidean group
\[
  \mathrm{E}(3) := \R^3\rtimes\mathrm{O}(3).
\]
Taking the reference stabilizer to be $\mathrm{O}(3)\hookrightarrow \mathrm{E}(3)$, we have the base space $\mathrm{E}(3)/\mathrm{O}(3)\cong\R^3$. Sites are then points in $\R^3$, with local structural symmetry groups given by subgroups of $\mathrm{O}(3)$. This is the geometry we will use to treat molecular aggregates. For the treatment of half-integer spin states, $\mathrm{Pin}(3)$ may be used in place of $\mathrm{O}(3)$.
\end{example}

\subsection{The site action and transport cocycle}
The quotient map \(q:G\to G/H\) organizes frames by their site positions: the fiber \(q^{-1}(x)\) is the set of frames above the position \(x\). This is a principal \(H\)-bundle, meaning that the right action of \(H\) relates any two frames in a fiber uniquely. When \(G\) is a Lie group and \(H\) is closed, the quotient \(G\to G/H\) additionally carries the structure of a smooth principal \(H\)-bundle. Its structure is expressed by the diagram\cite[Section~2.1 and~4.2]{husemoller94_fibre_bundl}
\begin{equation}
\label{eq:princ-bundle}
  \begin{tikzcd}[column sep=large]
    G\times H
    \arrow[r,shift left=0.6ex,"\gamma"]
    \arrow[r,shift right=0.6ex,swap,"\pi_{G}"]
    &G\arrow[r,"q"]&G/H
  \end{tikzcd}
\end{equation}
with right $H$-action \(\gamma(g, h) = gh\) and $G$-projection \(\pi_{G}(g, h) = g\). The identity \(q\circ\gamma=q\circ\pi_{G}\) expresses the invariance of \(q\) under the right \(H\)-action, whose orbits are the fibers of \(q\), each forming a right \(H\)-torsor (a set with a free and transitive right \(H\)-action). For the finite set \(X\subset G/H\), we choose one frame at each site. This choice is a section \(s:X\to G\), defined as a right inverse of \(q\) over \(X\), namely
\[
q\circ s=\iota_X,
\]
where \(\iota_X:X\hookrightarrow G/H\) denotes the inclusion. In our construction, the choice of representatives (or frames) \(s(x)\) for each \(x\in X\) specifies the relative orientation of the sites. Different choices of frame at a fixed site are related by right multiplication by an element of \(H\). In the \(\mathrm{E}(3)\) case, each frame specifies a site position in \(\mathbb R^3\) and an orientation determined by an orthogonal transformation in \(\mathrm O(3)\).

Now define the setwise stabilizer
\[
  T_0 := \operatorname{Stab}_{G}(X) := \{g \in G \, \mid \, gX = X\}.
\]
For each $g\in T_0$, define its permutation of sites by
\[
  \sigma_g(x):=g\cdot x.
\]
The section \(s\) defines the transport cocycle, an \(H\)-valued map recording the local transformation needed to compare a transformed frame with the chosen frame at its destination,
\begin{equation}
\label{eq:transport-cocycle-definition}
  \kappa:T_0\times X\longrightarrow H,
  \qquad
  \kappa(g,x):=s(g\cdot x)^{-1}gs(x),
\end{equation}
which satisfies the cocycle identity
\begin{equation}
\label{eq:cocycle-identity}
  \kappa(g_1g_2,x)
  =
  \kappa(g_1,\sigma_{g_2}(x))\kappa(g_2,x).
\end{equation}
Rearranging Eq.~\ref{eq:transport-cocycle-definition} relates the action of \(T_0\) on the site frames to the induced permutation of sites:
\[
gs(x)=s(\sigma_g(x))\kappa(g,x).
\]

To distinguish site types, equip $X$ with a labeling map $l:X\to L$. The label-preserving setwise stabilizer is
\[
  T_0
  :=\{g\in G\mid gX=X,\;l\circ\sigma_g=l\}.
\]
Thus only sites carrying the same label may be exchanged. The transport cocycle and its identity restrict to this subgroup.
\subsection{Global symmetry from local symmetries}
Each site's internal symmetry is specified by its associated symmetry group $P_x\leq H$. In the global setting, the frame $s(x)$ embeds $P_x$ into $G$ as
\[
\widetilde P_x:=s(x)P_xs(x)^{-1}
\leq\operatorname{Stab}_G(x).
\]
The setwise stabilizer $T_0$ preserves only the site positions, not the embedded symmetries $\{\widetilde{P}_x\}$. This motivates the following definition.
\begin{definition}[Global symmetry group]
The global symmetry group $T\leq G$ is the maximal subgroup that preserves the finite site configuration $X\subset G/H$ together with its transported framed local-group data $\{s(x)P_x\}_{x\in X}$.
\end{definition}
\begin{theorem}
\label{thm:glob-sym}
Define the following conditions:
\begin{align*}
&\text{Condition A:}\qquad
&&\kappa(g,x)P_x\kappa(g,x)^{-1}=P_{\sigma_g(x)},\\
&\text{Condition B:}\qquad
&&\kappa(g,x) \in P_{\sigma_g(x)}.
\end{align*}
The global symmetry group is given by
\begin{equation}
\label{eq:global-group}
T=\left\{
 g\in T_0 \,\middle|\,
 g\text{ satisfies Conditions A and B for every }x\in X
\right\}.
\end{equation}
\end{theorem}

The group property and maximality statement are proved in
Appendix~\ref{app:group}.
Condition A in Theorem~\ref{thm:glob-sym} ensures that each local structural
symmetry group is transported to the prescribed group at its
destination site.
Condition B in Theorem~\ref{thm:glob-sym} ensures that site frames are
preserved modulo local structural symmetries.

\begin{example}
  Appendix~\ref{app:T-recipe} gives a recipe for constructing $T$ in the Euclidean setting for finite point groups $\{P_x \leq \mathrm{O}(3)\}_{x \in X}$ associated with chosen frames:  $s(X) \subset \mathrm{E}(3)$. As an example, $T$ is then calculated for the geometry presented in Figure~\ref{fig:framed-point-groups} and shown to be $C_{4h}$.
\end{example}

With $G$, $H$, and the site labels understood, the framed
local-symmetry datum is summarized by the record
\[
  \mathcal D:=\bigl(X,\{(P_x,s(x)P_x)\}_{x\in X}\bigr),
\]
from which the global symmetry group is constructed.

\section{Representation Theory}
\label{sec:representation}
The previous section developed a geometric structure relating local and global symmetries. The aim of this section is to construct a representation of the global symmetry group $T$ that carries the local symmetry data covariantly according to that geometry. The key result is expressed in the following theorem, the proof of which uses the machinery of induced representations.\cite[Section~7.6]{Berndt2007} However, its geometric input differs from the usual homogeneous-space setting. Rather than inducing directly from the reference stabilizer \(H\) to the ambient group \(G\), we begin with the finite site configuration of the preceding section, equipped with potentially distinct local structural symmetry groups $P_x\leq H$. Their relative frames determine the compatible global group \(T\), and the local representations are then assembled by induction from its aggregate site stabilizers, orbit by orbit, as will be shown.
\begin{theorem}[Local-to-global representation]
  \label{thm:frame-independence}
Given a representation $(\rho,K)$ of the reference stabilizer $H$, the framed local-symmetry datum $\mathcal D$ determines a vector space $W$ and a representation of the global symmetry group $T$,
\[
    \rho_{\mathrm{glob}}:T\longrightarrow\operatorname{Aut}(W),
\]
whose coordinate realizations are equivalent under arbitrary changes of fiber-coordinate identification $\phi_{s(x)}\mapsto\phi_{s(x)h_x}$, with $h_x\in H$.
\end{theorem}
\subsection{Local-to-global representation construction}
\label{sec:loc-to-glob-construction}
We begin with an \(H\)-representation \((\rho,K)\) and form the associated site fibers of the principal \(H\)-bundle in Eq.~\eqref{eq:princ-bundle}. Intuitively, this attaches a ``copy'' of the model representation space \(K\) to each site. Precisely, for each $x\in X$, define the associated fiber
\[
K_x:=(q^{-1}(x)\times K)/H,
\]
where the quotient is taken with respect to the equivalence relation
\[
(g,k)\sim(gh,\rho(h)^{-1}k),
\qquad h\in H.
\]
This is the standard associated-fiber construction.\cite[Section~7.6, pp.~162--163]{Berndt2007}
Each $a\in q^{-1}(x)$ determines a fiber-coordinate identification
\[
\phi_a:K\longrightarrow K_x,\qquad
k\longmapsto[a,k].
\]
The geometric section \(s\) selects one such fiber-coordinate identification for every site,
\[
\phi_{s(x)}:K\overset{\cong}{\longrightarrow}K_x.
\]

The geometric datum $\mathcal D$ fixes the local positions and orientations
modulo the local structural symmetries, and hence determines $T$. We distinguish these
geometric frames from the fiber-coordinate identifications used to describe
states in $K_x$. With $\mathcal D$ fixed, replacing
\[
  \phi_{s(x)}\longmapsto\phi_{s(x)h_x}
  =\phi_{s(x)}\circ\rho(h_x),
  \qquad h_x\in H,
\]
changes only the coordinate identification. Such changes of fiber-coordinate identification are gauge changes
analogous to changes of body-frame convention in
molecular mechanics.\cite[p.~229]{littlejohn97_gauge_field_separ_rotat_andin}
Note, the elements $s(x)h_x$ here index alternative fiber-coordinate identifications and do not replace
the geometric frames defining $\mathcal D$. The fiber-coordinate independence asserted in
Theorem~\ref{thm:frame-independence} is proved in
Appendix~\ref{app:frames}.

Having identified the local fibers, the remaining task is to determine how the global symmetry group \(T\) acts upon them. How $T$ permutes sites is captured by decomposing the finite site configuration into $T$-orbits,
\[
  X=\bigsqcup_{i=1}^{m}\mathcal O_i,
  \qquad
  \mathcal O_i=T\cdot x_i,
\]
where $x_i$ is a representative of the $i$th orbit. Each orbit is naturally identified with the corresponding coset space of its stabilizer,
\[
  \mathcal O_i
  \cong
  T/T_{x_i},
\]
where
\begin{equation}
\label{eq:site-stab}
  T_{x_i}
  =
  \operatorname{Stab}_{T}(x_i)
\end{equation}
is the aggregate site stabilizer. The geometric significance of this orbit decomposition is provided by Theorem~\ref{thm:glob-sym}. If $x,y\in\mathcal O_i$, then there exists $g\in T$ such that
\[
  y=\sigma_g(x),
\]
and the corresponding local structural symmetry groups satisfy
\[
  P_w
  =
  \kappa(g,x)P_x\kappa(g,x)^{-1}.
\]
Thus every site within an orbit carries the same local symmetry data, transported covariantly by the cocycle.

For each orbit representative $x_i$, restrict the cocycle to the corresponding aggregate site stabilizer,
\begin{equation}
\label{eq:stabilizer-cocycle-restriction}
  \kappa_{x_i}:=\kappa(\,\cdot\,,x_i)\vert_{T_{x_i}}
  :T_{x_i}\longrightarrow P_{x_i},
\end{equation}
and define the local stabilizer representation
\begin{equation}
\label{eq:local-stab-rep}
  \pi_{x_i}
  :=\rho\circ\kappa_{x_i}
  :T_{x_i}\longrightarrow\operatorname{Aut}(K).
\end{equation}
The representation $\pi_{x_i}$ describes how the local fiber transforms under those global symmetries fixing the representative site $x_i$. It therefore provides the local datum from which the representation on the entire orbit is assembled.

Define the representation space associated with each orbit by
\[
  W_{\mathcal O_i}:=\bigoplus_{x\in\mathcal O_i}K_x.
\]
Each orbit consists of sites connected by global symmetries, including site permutations, while distinct orbits remain invariant. Under the transport action defined below, the space associated
with each orbit carries an induced representation
(see \cites[Section~7.6, pp.~161--162]{Berndt2007}
      [Section~7.1, Proposition~19, p.~55]{serre77_linear_repres_finit_group}):
\[
  W_{\mathcal O_i}
  \cong
  \operatorname{Ind}_{T_{x_i}}^{T}\pi_{x_i}.
\]
The complete local-to-global representation space is therefore
\[
  W:=\bigoplus_{i=1}^{m}W_{\mathcal O_i}
  =\bigoplus_{x\in X}K_x
  \cong
  \bigoplus_{i=1}^{m}
  \operatorname{Ind}_{T_{x_i}}^{T}\pi_{x_i}.
\]
This orbitwise assembly has a crystallographic counterpart in the construction of band representations from site-stabilizer representations.\cite{cano21_band_repres_topol_quant_chemis}

For each site $x\in X$, let
\begin{equation}
  \label{eq:incl-proj}
  \iota_x:K_x\longrightarrow W,
  \qquad p_x:W\longrightarrow K_x
\end{equation}
denote the canonical inclusion and projection. For $g\in T$, define
the transport map
\[
  \tau_x(g):K_x\longrightarrow K_{\sigma_g(x)},
  \qquad [a,k]\longmapsto[ga,k].
\]
This map is well-defined because left multiplication by $g$ commutes with
the right $H$-action defining the associated fibers. The global action is
\begin{equation}
\label{eq:global-block-action}
  \rho_{\mathrm{glob}}(g)
  :=\sum_{x\in X}\iota_{\sigma_g(x)}\tau_x(g)p_x,
  \qquad g\in T,
\end{equation}
which transports each site component and includes it in the target fiber.

\begin{proposition}
Eq.~\eqref{eq:global-block-action} defines a representation of
\(T\) on \(W\).
\end{proposition}
\begin{proof}
For $g_1,g_2\in T$,
\[
  \tau_{\sigma_{g_2}(x)}(g_1)\circ\tau_x(g_2)
  =\tau_x(g_1g_2),
  \qquad \tau_x(e)=I_{K_x},
\]
since successive transport sends $[a,k]$ to $[g_1g_2a,k]$.
The canonical projections and inclusions therefore give
$\rho_{\mathrm{glob}}(g_1)\rho_{\mathrm{glob}}(g_2)
=\rho_{\mathrm{glob}}(g_1g_2)$ and $\rho_{\mathrm{glob}}(e)=I_W$.
\end{proof}

Using the fiber-coordinate identifications $\phi_{s(x)}$, the identity
$gs(x)=s(\sigma_g(x))\kappa(g,x)$ gives
\begin{equation}
\label{eq:transport-coordinates}
  \tau_x(g)
  =\phi_{s(\sigma_g(x))}\rho\bigl(\kappa(g,x)\bigr)\phi_{s(x)}^{-1}.
\end{equation}
The relation between the intrinsic global action and its local-coordinate
realization may be expressed by the commutative diagram
\[
\begin{tikzcd}[column sep=large,row sep=large]
K \arrow[r,"{\rho(\kappa(g,x))}"]
  \arrow[d,"\phi_{s(x)}"']
& K \arrow[d,"\phi_{s(\sigma_g(x))}"] \\
K_x \arrow[r,"{\tau_x(g)}"]
    \arrow[d,"\iota_x"']
& K_{\sigma_g(x)} \arrow[d,"\iota_{\sigma_g(x)}"] \\
W \arrow[r,"\rho_{\mathrm{glob}}(g)"']
& W
\end{tikzcd}
\]
where the cocycle $\kappa$ describes transport in these fiber coordinates.
Different fiber-coordinate identifications give equivalent coordinate
realizations of the same intrinsic representation.

\paragraph{Notation.} For site-resolved operators and subspaces, we use a tilde for their realizations in the model space $K$, no decoration for their realizations in a site fiber $K_x$, and an overline for their embeddings in $W$. For a model operator $\widetilde A_x\in\operatorname{End}(K)$ and a model subspace $\widetilde L_x\subseteq K$, these conventions mean

\begin{equation}
\label{eq:hat-convention}
\begin{aligned}
  A_x&:=\phi_{s(x)}\widetilde A_x\phi_{s(x)}^{-1},
  &\qquad \overline A_x&:=\iota_xA_xp_x,\\
  L_x&:=\phi_{s(x)}\widetilde L_x\subseteq K_x,
  &\qquad \overline L_x&:=\iota_xL_x\subseteq W.
\end{aligned}
\end{equation}

\subsection{Relation to global quantum representations}
The construction of the preceding subsection begins with unitary representations describing the quantum states of individual sites. Together with the geometry of Section~\ref{sec:geometry}, these local representations determine a state space \(W\) carrying a covariant action of the aggregate symmetry group \(T\). Although constructed from physically interpretable local data, \(W\) is, at this stage, an abstract representation. Its relationship to the physical quantum state space of the aggregate remains to be established. To this end, we now construct an intertwining map (see \cite[Section~1.2, Definition~1.4, p.~9]{Berndt2007}) that relates the local-to-global representation to the quantum representation of the ambient transformation group. Let
\[
U:G\longrightarrow\mathrm{U}(\mathcal H)
\]
be a unitary representation on a physical Hilbert space \(\mathcal H\).
\begin{example}
  \label{exp:E3-uni-act}
  For $G=\mathrm{E}(3)$, the physical Hilbert space of spatial
  wavefunctions is $\mathcal H=L^2(\R^3)$, with unitary action
  \[
    (U(g)\Psi)(x):=\Psi(g^{-1}\cdot x).
  \]
\end{example}

Now suppose that the local $H$-representation $(\rho,K)$ admits an $H$-equivariant map
\[
  j:K\longrightarrow\mathcal H,
  \qquad
  j\circ\rho(h)=U(h)\circ j,
  \quad h\in H.
\]
For a vector $\psi=(\psi_x)_{x\in X}\in W$, let
\[
  f_x:=\phi_{s(x)}^{-1}(\psi_x)\in K
\]
be its model-space coordinate in the chosen frame. Define the reconstruction
map
\[
  F_s:W\longrightarrow\mathcal H,
  \qquad
  F_s(\psi):=\sum_{x\in X}U(s(x))j(f_x).
\]
\begin{proposition}[Physical realization]
\label{prop:reconstruct-intertwiner}
The reconstruction map $F_s$ intertwines the local-to-global representation
with the physical representation $U\vert_{T}$:
\[
  F_s\circ\rho_{\mathrm{glob}}(g)
  =U(g)\circ F_s,
  \qquad g\in T.
\]
\end{proposition}

\begin{proof}
For $g\in T$ and $\psi\in W$,
\begin{align*}
  U(g)F_s(\psi)
  &=\sum_{x\in X}U(gs(x))j(f_x) \\
  &=\sum_{x\in X}U(s(\sigma_g(x)))
    U(\kappa(g,x))j(f_x) \\
  &=\sum_{x\in X}U(s(\sigma_g(x)))
    j\bigl(\rho(\kappa(g,x))f_x\bigr) \\
  &=F_s\bigl(\rho_{\mathrm{glob}}(g)\psi\bigr).
\end{align*}
The third equality uses the $H$-equivariance of $j$, while the final equality
follows from Eq.~\eqref{eq:global-block-action}
and~\eqref{eq:transport-coordinates}.
\end{proof}
No injectivity or surjectivity of $F_s$ is implied without additional assumptions on $j$ and the transported local
subspaces.

For fixed $\mathcal D$, replace $\phi_{s(x)}$ by $\phi_{s(x)b_x}$,
with $b_x\in H$. In block coordinates, define the change-of-coordinates
operator $B$ by
\[
B:=\bigoplus_{x\in X}\rho(b_x).
\]
Then
\[
F'=F_sB,
\qquad
\rho_{\mathrm{glob}}'(g)=B^{-1}\rho_{\mathrm{glob}}(g)B,
\quad g\in T.
\]
Here $F'$ and $\rho_{\mathrm{glob}}'$ denote the reconstruction map and
global action in the new fiber coordinates. Consequently, the reconstructed
physical subspace is independent of the fiber-coordinate identifications,
while the geometric datum $\mathcal D$ and its global group $T$ remain fixed.

Through the reconstruction map \(F_s\), a collection of site states \(\psi=(\psi_x)_{x\in X}\in W\), each belonging to a framed copy \(K_x\) of the local model space \(K\), is mapped into the physical Hilbert space \(\mathcal H\)--combining to form an aggregate quantum state. In \(\mathcal H\), these contributions share a common physical inner product that accounts for their mutual overlaps. Although the site fibers are orthogonal by construction in \(W\), their physical realizations in \(\mathcal H\) need not be. The intertwining property of Proposition~\ref{prop:reconstruct-intertwiner} guarantees that this reconstruction respects the aggregate symmetry, i.e., transforming the local site data under \(\rho_{\mathrm{glob}}\) and then reconstructing gives the same physical state as reconstructing first and applying \(U\).

A benefit of the reconstruction map is that the local model space \(K\)
may retain only reduced degrees of freedom relevant to the symmetry under
consideration. For example, consider electron transfer between sites whose
local environments have different orientations. Models incorporating
spin–orbit coupling can retain local orbital and spin degrees of freedom
while absorbing radial dependence into effective coupling parameters.
Relating these local descriptions requires accounting for their positions
and orientations within the aggregate. The following example establishes
the orbital part of this construction by fixing a radial profile and
reconstructing local angular states in the full spatial Hilbert space.
Explicitly treating spin–orbit coupling would additionally require
incorporating spinor representations.

\begin{example}[Reduced angular model]
\label{exp:Reduced-angular-model}
Take $G=\mathrm{E}(3)$ and $H=\mathrm{O}(3)$, with the unitary action given in Example~\ref{exp:E3-uni-act}.
Using spherical coordinates,
\[
  L^2(\R^3)
  \cong L^2(\R^+,r^2dr)\otimes L^2(S^2).
\]
To retain only the angular degrees of freedom, take $K=L^2(S^2)$ with
\[
  (\rho(R)f)(\Omega)=f(R^{-1}\Omega),
  \qquad R\in \mathrm{O}(3).
\]
Fix a nonzero radial profile $R_0\in L^2(\R^+,r^2dr)$ and define the
$\mathrm{O}(3)$-equivariant map
\[
  j:L^2(S^2)\longrightarrow L^2(\R^3),
  \qquad
  (jf)(r,\Omega):=R_0(r)f(\Omega).
\]
Rotations act only on the angular coordinate, so
$j\rho(R)=U(R)j$. For $\psi=(f_x)_{x\in X}$ in fiber coordinates, the
reconstruction is
\[
  F_s(\psi)=\sum_{x\in X}U(s(x))j(f_x).
\]
\end{example}
Conventionally, expressing angular states localized at different sites in a common coordinate system can involve rotation formulas and translation addition theorems for solid spherical harmonics.\cite{STEINBORN19731} The local-to-global representation instead provides a symmetry-covariant description of these states directly in terms of their local representations. Consequently, symmetry constraints on intersite couplings can be determined without explicitly performing such basis transformations. This is particularly useful for interactions involving reduced degrees of freedom such as having only angular coordinates, as with spin–orbit coupling.

\section{Local and global sectors and their compatibility}
\label{sec:sector-compatibility}
We now compare the local symmetry decomposition of the site representations $(\pi_x, K_x)$ with the global isotypic decomposition of the local-to-global representation $(\rho_{\mathrm{glob}},W)$. For simplicity, we assume throughout this section that the local structural symmetry groups $\{P_x\}$ are finite and that the representations are unitary.

\subsection{The isotypic decomposition}
\label{sec:isotypic-decomp}
A standard consequence of Schur's lemma (see~\cites[Theorem~1.2]{Berndt2007}[Section~2.2, Proposition~7]{serre77_linear_repres_finit_group}) is that the isotypic components (see~\cite[Section~7.2, Proposition~19]{serre77_linear_repres_finit_group}) of a $G$-representation are invariant under every $G$-equivariant endomorphism. In particular, they define invariant subspaces for quantum dynamics generated by a $G$-symmetric Hamiltonian. We first recall this general result using a multiplicity-space construction, then examine its consequences for dynamics in the local-to-global representation.

Let $(\pi,W)$ be a unitary representation of a finite group $G$, and let $\Irr(G)$ denote the equivalence classes of irreducible $G$-representations. For each $\lambda\in\Irr(G)$, choose a representative $(\pi_\lambda,V_\lambda)$ and define the multiplicity space
\[
  M_\lambda:=\operatorname{Hom}_{G}(V_\lambda,W).
\]
The corresponding $\lambda$-isotypic component is
\[
  W_\lambda
  :=
  \operatorname{ev}_\lambda
  (M_\lambda\otimes V_\lambda),
\]
where the evaluation map is the isomorphism
\[
  \operatorname{ev}_\lambda:
  M_\lambda\otimes V_\lambda\xlongrightarrow{\cong} W_\lambda,
  \qquad
  \phi\otimes x\longmapsto\phi(x).
\]
The isotypic decomposition of $(\pi,W)$ is then
\[
  W=\bigoplus_{\lambda\in\Irr(G)}W_\lambda.
\]

For the character $\chi_\lambda$ and dimension $d_\lambda:=\dim V_\lambda$, the orthogonal projector onto $W_\lambda$ is given by the finite character sum
(see \cite[Section~2.3, Theorem~2.3]{Berndt2007}):
\[
  \operatorname{Proj}_\lambda^{G}(\pi,W)
  :=
  \frac{d_\lambda}{|G|}
  \sum_{g\in G}
  \chi_\lambda(g)^*\pi(g).
\]
For compact groups, the corresponding expression is obtained by replacing the normalized finite sum with integration against the normalized Haar measure.\cite[Section~3.3, Theorem~3.4 and Proposition~3.2,
pp.~39--40]{Berndt2007}

We now show that the isotypic components of $W$ are invariant under any $G$-intertwining operator $A\in\operatorname{End}_{G}(W)$, which by definition satisfies
\[
  A \circ \pi(g)=\pi(g) \circ A,
  \qquad g\in G
\]
For any $\phi\in M_\lambda$, equivariance gives
\[
  A\circ\phi\circ\pi_\lambda(g)
  =
  A\circ\pi(g)\circ\phi
  =
  \pi(g)\circ A\circ\phi,
\]
and hence $A\circ\phi\in M_\lambda$. Composition with $A$ therefore defines an operator on the multiplicity space,
\[
  A_\lambda:M_\lambda\longrightarrow M_\lambda,
  \qquad
  \phi\longmapsto A\circ\phi.
\]
By construction,
\[
  A\circ\operatorname{ev}_\lambda
  =
  \operatorname{ev}_\lambda
  \circ(A_\lambda\otimes I_{V_\lambda}).
\]
Since $\operatorname{ev}_\lambda$ is an isomorphism onto
$W_\lambda$, it follows that
\[
  A|_{W_\lambda}
  =
  \operatorname{ev}_\lambda
  (A_\lambda\otimes I_{V_\lambda})
  \operatorname{ev}_\lambda^{-1}.
\]
In particular,
\[
  A(W_\lambda)\subseteq W_\lambda,
  \qquad
  \forall\lambda\in\Irr(G).
\]
Thus every intertwining operator preserves the isotypic decomposition, acting independently on the multiplicity spaces associated with each irreducible representation class. A quantum Hamiltonian, however, may not be bounded and is therefore not necessarily defined on all of $W$. Nevertheless, by Stone's theorem, a self-adjoint Hamiltonian $H$ generates a strongly continuous unitary group $U(t)=e^{-itH/\hbar}$ (see \cite[Sections~10.1--10.2.1, pp.~142--143]{woit17_quant_theor_group_repres}). If $H$ is $G$-equivariant, including invariance of its domain, then $U(t)$ is also $G$-equivariant. As a bounded intertwining operator, $U(t)$ therefore preserves the isotypic decomposition.

\subsection{Site-resolved local projectors and global projectors}
Motivating this work is the question of whether local symmetries impose constraints on the dynamics of the aggregate. The weakest answer in the affirmative is that the local symmetries and their geometric framing determine the global symmetry, while the constructed global representation determines the isotypic decomposition of the aggregate state space. For an arbitrary state in the aggregate state space, without determining its compatibility with the local structural symmetries, we obtain only the constraints imposed by the global symmetry. To translate these global constraints into statements about local symmetry sectors, we must establish how local isotypic components are supported within the global isotypic decomposition.

To determine the local isotypic components of site $x$ in $W$, restrict the local model representation $(\rho,K)$ from $H$ to the local structural symmetry group $P_x$. For \(\lambda\in\Irr(P_x)\), define the local isotypic projector and its image in the model space by
\begin{equation}
\label{eq:local-isotypic-representation}
  \widetilde{\mathcal Q}_{x,\lambda}
  :=\operatorname{Proj}_{\lambda}^{P_x}(\rho|_{P_x},K),
  \qquad
  \widetilde K_{x,\lambda}
  :=\operatorname{Ran}\widetilde{\mathcal Q}_{x,\lambda}.
\end{equation}
Following the notation convention above (Eq.~\ref{eq:hat-convention}), their fiber realizations are denoted by \(\mathcal Q_{x,\lambda}\) and \(K_{x,\lambda}\), and their aggregate embeddings by \(\overline{\mathcal Q}_{x,\lambda}\) and \(\overline K_{x,\lambda}\). The latter is the site-resolved local isotypic sector \((x,\lambda)\) in \(W\).
We emphasize that the local projectors are defined using the full local structural symmetry group \(P_x\), not only the subgroup of local symmetries inherited from \(T\); a point that will be key in determining local-to-local symmetry exclusions imposed by the global symmetry.

Similarly, the global isotypic projectors of $(\rho_{\mathrm{glob}},W)$ are
\[
  \mathcal{P}_\gamma
  :=\operatorname{Proj}_\gamma^{T}
  (\rho_{\mathrm{glob}},W),
  \qquad \gamma\in\Irr(T).
\]

\subsection{Branching and global support}
We say that two Hilbert subspaces $A$ and $B$ have nonzero overlap if there exist $a\in A$ and $b\in B$ such that $\braket{a,b}\neq 0$. Equivalently, for their orthogonal projectors $\mathcal{P}_{A}$ and $\mathcal{P}_{B}$, nonzero overlap means $\mathcal{P}_{A}\mathcal{P}_{B}\neq 0$.

Assuming a finite model space $(\rho, K)$, we define the local-to-global overlap coefficient as the squared Hilbert–Schmidt norm
\begin{equation}
\label{eq:local-global-overlap}
  b_{x,\lambda}^{\gamma}
  :=\operatorname{Tr}
  \left(\mathcal{P}_\gamma\overline{\mathcal{Q}}_{x,\lambda}\right) = \operatorname{Tr}
  \left((\mathcal{P}_\gamma\overline{\mathcal{Q}}_{x,\lambda})^*\mathcal{P}_\gamma\overline{\mathcal{Q}}_{x,\lambda}\right)
  =\left\|\mathcal{P}_\gamma\overline{\mathcal{Q}}_{x,\lambda}
  \right\|_{\mathrm{HS}}^2.
\end{equation}
Since $\mathcal{P}_\gamma$ and $\overline{\mathcal{Q}}_{x,\lambda}$ are each orthogonal projectors,
\[
  b_{x,\lambda}^{\gamma}\geq 0,
  \qquad
  b_{x,\lambda}^{\gamma}>0
  \iff \mathcal{P}_\gamma\overline{\mathcal{Q}}_{x,\lambda}\neq 0.
\]
Thus $b_{x,\lambda}^{\gamma}>0$ precisely when the local isotypic component $\overline K_{x,\lambda}$ and the global isotypic component $W_\gamma:=\operatorname{Ran}\mathcal{P}_\gamma$ have nonzero overlap. As such, $b_{x,\lambda}^\gamma$ quantifies the geometric alignment of these subspaces within the aggregate Hilbert space.

\subsection{Symmetry-adapted operators and dynamic compatibility}
Applying the general isotypic invariance result of Section~\ref{sec:isotypic-decomp} to \((\rho_{\mathrm{glob}},W)\), every \(T\)-equivariant operator preserves the global isotypic components \(W_\gamma\). In particular, a Hamiltonian invariant under \(\rho_{\mathrm{glob}}\) generates dynamics that cannot couple inequivalent global isotypic components. Consequently, if two local isotypic components do not overlap with a shared global component, then those local components will not mix through that global component. Quantitatively
\[
  b_{x_1, \lambda_1}^\gamma b_{x_2, \lambda_2}^\gamma = 0.
\]
If the local isotypic components overlap with no shared global component, such that
\[
  \sum_{\gamma \in \Irr(T)} b_{x_1, \lambda_1}^\gamma b_{x_2, \lambda_2}^\gamma = 0,
\]
then they cannot mix under the \(T\)-symmetric global dynamics.

By Proposition~\ref{prop:reconstruct-intertwiner}, whenever an \(H\)-equivariant map \(j:K\to\mathcal H\) exists, the local-to-global representation intertwines with the physical representation \(U|_{T}\). The condition of global equivariance can also be expressed in terms of site blocks. For \(\overline A \in\operatorname{End}(W)\), define
\[
A_{wx}:=p_w\overline A\iota_x:K_x\longrightarrow K_w
\]
and
\[
\widetilde A_{wx}:= \phi_{s(w)}^{-1}A_{wx}\phi_{s(x)} \in \operatorname{End}(K)
\]
Then \(\overline A\) commutes with \(\rho_{\mathrm{glob}}(T)\) if and only if
\[
\widetilde A_{\sigma_g(w),\sigma_g(x)}
=
\rho\bigl(\kappa(g,w)\bigr)
\widetilde A_{wx}
\rho\bigl(\kappa(g,x)\bigr)^{-1}
\]
for all \(g\in T\) and \(x,w\in X\). Thus, global symmetry relates operator blocks by simultaneously transforming their source and target sites.

\section{Application}
\label{sec:application}
The local-to-global symmetry framework is now sufficiently developed for some practical applications, the first of which is the construction of a symmetry-adapted Hamiltonian within a site basis. Conventionally, site-basis Hamiltonians are assembled from local Hamiltonians describing the individual sites and coupling operators describing their interactions, with matrix elements obtained from microscopic calculations or introduced phenomenologically.\cite{hestand18_expan_theor_h_j_molec_aggreg,KUHN201147} Symmetry-based approaches can further classify states according to irreducible representations and establish selection rules for their interactions.\cite{wigner2013group,eckart30_applic_group_theor_to_quant,Cotton1990-td} Here, we use the local-to-global representation to construct operators
invariant under the global symmetry while retaining the geometric and local
symmetry structure of the site basis. We then specialize to self-adjoint
Hamiltonians and relate the local isotypic sectors to the global Hamiltonian
blocks to resolve coupling constraints and dynamics by site. This requires the framed representation as well as the abstract symmetry groups.
\subsection{Operator symmetrization}
\label{sec:operator-symmetrization}
An arbitrary operator $A\in\operatorname{End}(W)$ can be projected onto the commutant by Reynolds averaging
(see \cite[Section~3.1.2, p.~73]{derksen15_comput_invar_theor}):
\begin{equation}
  \mathcal R_{T}(A)
  :=\frac{1}{|T|}
  \sum_{g\in T}
  \rho_{\mathrm{glob}}(g)A\rho_{\mathrm{glob}}(g)^\dagger.
\end{equation}
This map satisfies
\[
  \mathcal R_{T}^2=\mathcal R_{T},
  \qquad
  \operatorname{im}\mathcal R_{T}
  =\operatorname{End}_{T}(W),
\]
and is the orthogonal projector onto the commutant with respect to the Hilbert--Schmidt inner product (Appendix~\ref{app:reynolds-projection}). It also preserves adjoints, so $\mathcal R_{T}(A)$ is a symmetry-compatible Hamiltonian whenever $A=A^\dagger$. For compact $T$, the normalized Haar integral gives the corresponding projection.
\begin{figure}[h]
\centering
\includegraphics[width=0.65\linewidth]{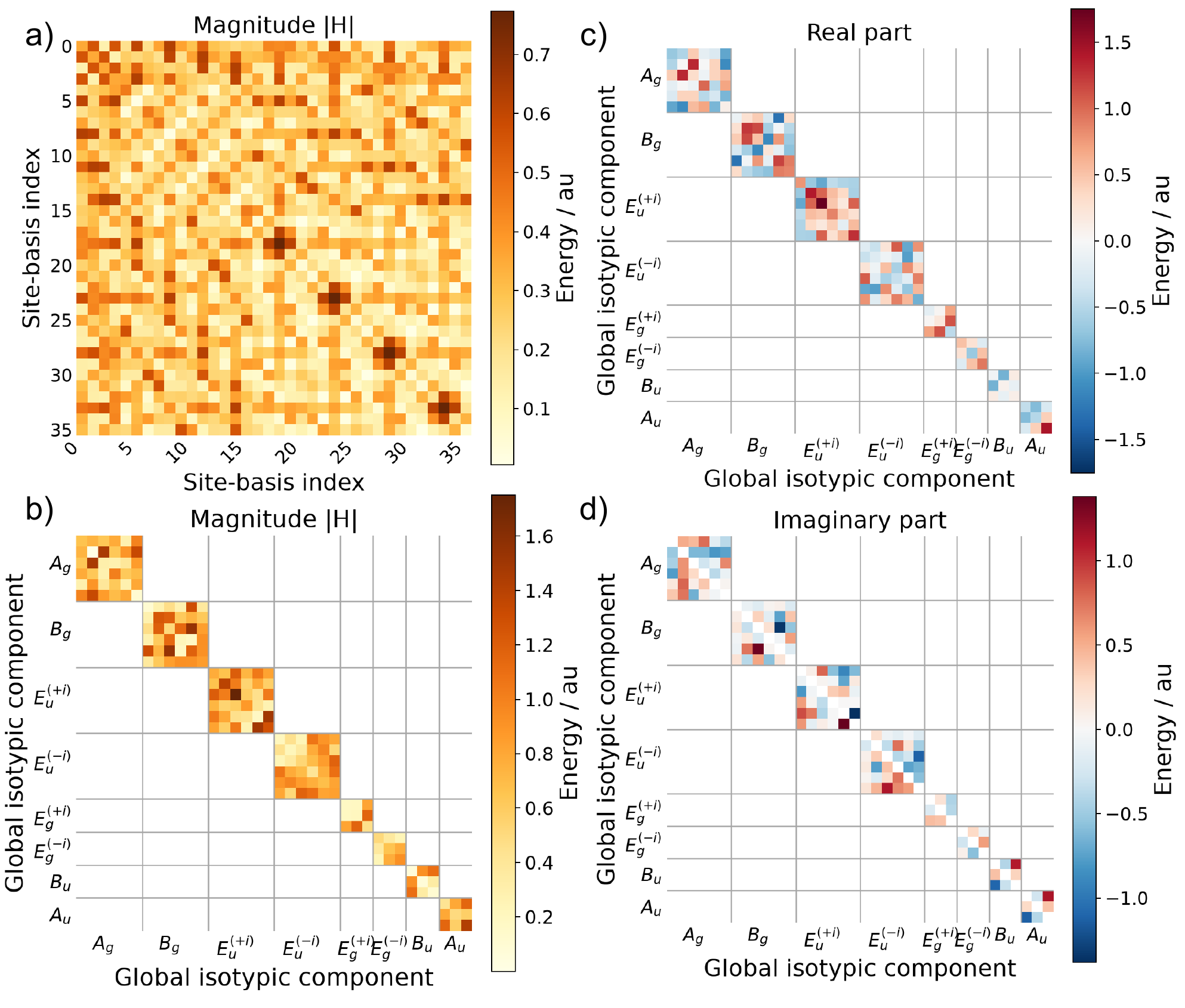}
\caption{Visualization of a Hermitian Hamiltonian with randomized non-zero elements, constructed in a site basis of the square molecular geometry depicted in Figure~\ref{fig:framed-point-groups} and symmetrized according to the local-to-global symmetry construction using Reynolds averaging. The Wigner D-matrix representation, truncated to $l_\text{max} = 2$, provided the local model representation $(\rho, K)$, expanded in a basis of spherical harmonics. (a) Matrix-element
magnitudes $|H_{ij}|$ of the symmetrized Hamiltonian in the spherical harmonics site basis. (b) Matrix-element magnitudes, (c) real parts, and (d) imaginary parts, in the symmetry-adapted basis, revealing the block structure and vanishing interblock couplings. Color scales indicate energy in atomic units.}
\label{fig:square-hamiltonian}
\end{figure}

We demonstrate the operator symmetrization for a Hamiltonian constructed in a basis of four sites, with each site having a basis of spherical Harmonics truncated to $l_\text{max} = 2$. The Hamiltonian was constructed from upper-triangular complex matrix \(A\), with elements randomly sampled from an annulus in the complex plane, and a real diagonal random matrix \(D\) with elements sampled from the intersection of annulus and the real line. The Hamiltonian was then taken as $H = D + A + A^\dagger$.  This was done to give a general Hamiltonian form with non-zero intersite and intrasite couplings so that upon symmetrization, the block structure of the Hamiltonian can be attributed fully to the symmetry constraints arising from geometric structure.

Prior to specifying the framed geometry, \(H\) is an operator on an abstract direct sum of site spaces,
without a prescribed realization in a common ambient geometry.
Such a realization requires identifying how the local states transform
within that geometry and reconstruct into a common physical state space.
The local-to-global construction supplies these identifications, while symmetrization enforces compatibility of the Hamiltonian with the resulting aggregate symmetry. To this end, the four site system is imbued with the geometry of the square site configuration shown in Figure \ref{fig:framed-point-groups}, where each site has a local $C_{2v}$ symmetry and shares the same label. According to Example~\ref{exp:E3-quotient-structure}, $H = \mathrm{O}(3)$. The Wigner D-matrix representation\cite{wigner2013group} was taken as the model local $\mathrm{O}(3)$-representation, from which a global $T$-representation was constructed from the $\mathcal D$ datum of the square site geometry (see Appendix~\ref{app:T-recipe} for details). Figure~\ref{fig:square-hamiltonian} illustrates the resulting symmetry-adapted block structure for the square aggregate. Figure~\ref{fig:square-hamiltonian}a shows the magnitudes of the symmetrized Hamiltonian in the site basis. No block-diagonal structure is apparent in this basis. Figure~\ref{fig:square-hamiltonian}b--d show the Hamiltonian in a symmetry-adapted basis illustrating the resulting block structure for the square-aggregate system. Each block corresponds to an isotypic block of the global representation, with the axes giving the corresponding $C_{4h}$ irrep labels. The labels $E_{g/u}^{(+i)}$ and $E_{g/u}^{(-i)}$ distinguish the complex-conjugate one-dimensional irreps whose characters on a chosen $C_4$ generator are $+i$ and $-i$, respectively.

\subsection{Hamiltonian coupling channels}
The global block structure identifies components that cannot mix under the Hamiltonian. Our interest, however, is in states prepared and observed at particular sites. Consider a state initially localized at an isolated site, whose dynamics are constrained by the site's local structural symmetry. Such a state may be specified by its membership in a local isotypic component, labeled by an irreducible representation \(\lambda_{\mathrm{initial}}\), or a direct sum of such components. Upon introducing the aggregate environment, the global isotypic components become the dynamically invariant subspaces. As illustrated in Fig. \ref{fig:square-orbitals}b, we wish to determine whether an initially prepared local symmetry sector can evolve into another local sector labeled by \(\lambda_{\mathrm{final}}\). Importantly, local isotypic components need not be invariant under the global symmetry group \(T\), particularly when global transformations exchange sites. Consequently, these subspaces do not generally carry representations of \(T\) and cannot themselves be decomposed into global irreducible representations. Yet, they do have support in the global isotypic components. Rather, their embeddings in the aggregate state space, $W$, determine their overlap with the global isotypic components. The local-to-global representation relates these two decompositions and thereby establishes dynamical exclusions between local symmetry sectors. As in the preceding sections, all representations are assumed unitary. The Hamiltonian \(H\) is self-adjoint and assumed to commute with \(\rho_{\mathrm{glob}}(T)\).

A local decomposition makes it possible to ask which symmetry sectors at
one site couple to those at another. To retain the global origin of the
constraints, we resolve the Hamiltonian by both its global sectors and its
local endpoints. Define the channel from $(x_j,\mu)$ to $(x_i,\lambda)$ by
\begin{equation}
\label{eq:resolved-channel}
  H_{i\lambda,j\mu}^{\gamma,\gamma'}
  :=\overline{\mathcal{Q}}_{x_i,\lambda}\mathcal{P}_\gamma H \mathcal{P}_{\gamma'}
    \overline{\mathcal{Q}}_{x_j,\mu}.
\end{equation}
Here $\|A\|_{\mathrm{HS}}^2=\operatorname{Tr}(A^\dagger A)$.
The norm is independent of orthonormal basis choices within the endpoint
sectors and measures the strength of the corresponding operator block.

Global invariance first excludes channels between inequivalent global
isotypic components:
\[
  \mathcal{P}_\gamma H \mathcal{P}_{\gamma'}=0\qquad(\gamma\ne\gamma').
\]
For the remaining channels, local support gives the further implication
\begin{equation}
\label{eq:channel-support-exclusion}
  b_{x_i,\lambda}^{\gamma}b_{x_j,\mu}^{\gamma}=0
  \quad\Longrightarrow\quad \bigl\|H_{i\lambda,j\mu}^{\gamma,\gamma}\bigr\|_{\mathrm{HS}}=0.
\end{equation}
This resolves a global exclusion into statements about particular local
endpoints. The complete local Hamiltonian block is
\begin{equation}
\label{eq:total-local-channel}
  H_{i\lambda,j\mu}
  :=\overline{\mathcal{Q}}_{x_i,\lambda}H\overline{\mathcal{Q}}_{x_j,\mu}
  =\sum_\gamma H_{i\lambda,j\mu}^{\gamma,\gamma},
\end{equation}
the norm of which, i.e. $\|H_{i\lambda,j\mu}\|_{\mathrm{HS}}$, provides a basis-independent measure of the coupling strength between the two site isotypic components, after summing their globally resolved operator blocks. Note, it may vanish even when symmetry permits the channel.

\subsection{Predicting support by character scalar products}
The channel norms require a Hamiltonian, but many channel zeros can be
established before its matrix elements are specified. The required information
is the global support of each local endpoint. The induced representation above
allows this support to be calculated using characters on the aggregate site stabilizer.

Let $(\rho_{x,\lambda},\widetilde K_{x,\lambda})$ denote the \(\lambda\)-isotypic representation of \((\rho|_{P_x},K)\), on the subspace defined in Eq.~\eqref{eq:local-isotypic-representation}. Pulling this representation back along $\kappa_x:T_x\to P_x$ (Eq.~\ref{eq:stabilizer-cocycle-restriction}) gives the stabilizer representation:
\[
  \pi_{x,\lambda}:=\rho_{x,\lambda}\circ\kappa_x: T_x \to \operatorname{Aut}(\widetilde K_{x,\lambda}).
\]
Following the local-to-global construction in Section~\ref{sec:loc-to-glob-construction}, taking the direct sum of the transported local isotypic components over the orbit
$\mathcal O_x:=\{g\cdot x\mid g\in T\}\subseteq X$
gives a global invariant subspace:
\[
  W_{x,\lambda}
  :=\bigoplus_{w\in\mathcal O_x}
      \tau_x(g_w)K_{x,\lambda},
  \qquad g_w\cdot x=w.
\]
It carries the induced representation
\[
  W_{x,\lambda}
  \cong
  \operatorname{Ind}_{T_x}^{T}\pi_{x,\lambda}.
\]
The orbit construction therefore relates a site-local label to its
global symmetry content.

Let $\chi_{x,\lambda}$ denote the character of the stabilizer
representation $\pi_{x,\lambda}$ on $\widetilde K_{x,\lambda}$. For
$\gamma\in\Irr(T)$, Frobenius reciprocity gives
(see \cite[Section~7.2, Theorem~13]{serre77_linear_repres_finit_group}):
\begin{align}
\label{eq:channel-reciprocity}
  m_{x,\lambda}^{\gamma}
  &:=\left\langle\chi_\gamma,
      \chi_{\operatorname{Ind}_{T_x}^{T}
      \pi_{x,\lambda}}\right\rangle_{T}
  =\left\langle\chi_\gamma|_{T_x},
      \chi_{x,\lambda}\right\rangle_{T_x}\\
  &=\frac{1}{|T_x|}
    \sum_{h\in T_x}
    \overline{\chi_\gamma(h)}\,\chi_{x,\lambda}(h).
\nonumber
\end{align}
Here $m_{x,\lambda}^\gamma=\dim\operatorname{Hom}_T(V_\gamma,W_{x,\lambda})$ is the multiplicity of the irreducible $T$-representation $V_\gamma$ in the induced space $W_{x,\lambda}$. The global $\gamma$-isotypic component of $W$ is $W_\gamma=\operatorname{Ran}\mathcal P_\gamma$. The local-to-global overlap coefficient (Eq.~\ref{eq:local-global-overlap}) is proportional to this multiplicity. Expanding $\mathcal P_\gamma$ gives
\begin{align}
\label{eq:overlap-trace}
  b_{x,\lambda}^\gamma &:= \operatorname{Tr}(\mathcal{P}_\gamma\overline{\mathcal{Q}}_{x,\lambda})\\
\label{eq:overlap-character-trace}
  &= \frac{d_\gamma}{|T|}
    \sum_{g\in T}
    \overline{\chi_\gamma(g)}\,\operatorname{Tr}\big(
    \rho_{\mathrm{glob}}(g)\overline{\mathcal{Q}}_{x,\lambda}\big).
\end{align}
The projector $\overline{\mathcal Q}_{x,\lambda}$ vanishes outside the embedded site fiber $K_x$, while $\rho_{\mathrm{glob}}(g)$ sends that fiber into $K_{\sigma_g(x)}$. Thus the trace $\operatorname{Tr}\big(\rho_{\mathrm{glob}}(g)\overline{\mathcal Q}_{x,\lambda}\big)$ vanishes when $\sigma_g(x)\ne x$. Only elements $g\in T_x$ therefore contribute to the sum in Eq.~\ref{eq:overlap-character-trace}. For such elements, this trace equals the stabilizer character $\chi_{x,\lambda}(g)$, giving
\[
  b_{x,\lambda}^{\gamma}
  =\frac{d_\gamma}{|T|}
    \sum_{g\in T_x}
      \overline{\chi_\gamma(g)}\,\chi_{x,\lambda}(g).
\]
Using $|\mathcal O_x|=|T|/|T_x|$ we then have
\begin{equation}
\label{eq:overlap-reciprocity}
  b_{x,\lambda}^{\gamma}
  =\frac{d_\gamma}{|\mathcal O_x|}\,m_{x,\lambda}^{\gamma},
  \qquad d_\gamma=\dim V_\gamma.
\end{equation}
The multiplicity $m_{x,\lambda}^\gamma$ calculated by the character scalar product in Eq.~\eqref{eq:channel-reciprocity} provides a practical criterion for local-to-global exclusion: $m_{x,\lambda}^\gamma=0$ precisely when the local isotypic component has zero overlap with the global $\gamma$-isotypic component. Similarly, $m_{x_i,\lambda}^\gamma m_{x_j,\mu}^\gamma=0$ excludes the site-to-site mixing channel through $\gamma$. If this product vanishes for every $\gamma$, the two local sectors cannot mix under any globally invariant Hamiltonian.

We emphasize that the exclusion criterion can be evaluated without constructing the model representation $(\rho, K)$ explicitly. If $\widetilde K_{x,\lambda}$ contains $n_{x,\lambda}$ equivalent copies of the $P_x$-irrep $\lambda$, then
\[
  m_{x,\lambda}^{\gamma}
  =n_{x,\lambda}
  \left\langle
    \chi_\gamma|_{T_x},\chi_\lambda\circ\kappa_x
  \right\rangle_{T_x}.
\]
For any model containing this local sector, the multiplicity $n_{x,\lambda}>0$ does not affect whether the overlap vanishes. Thus the symmetry-enforced exclusion can be determined directly from the character tables and the framed stabilizer map
$\kappa_x:T_x\to P_x$.

\subsection{Square aggregate: coupling channels and dynamics}
\begin{figure}[htb]
\centering
\includegraphics[width=0.3\linewidth]{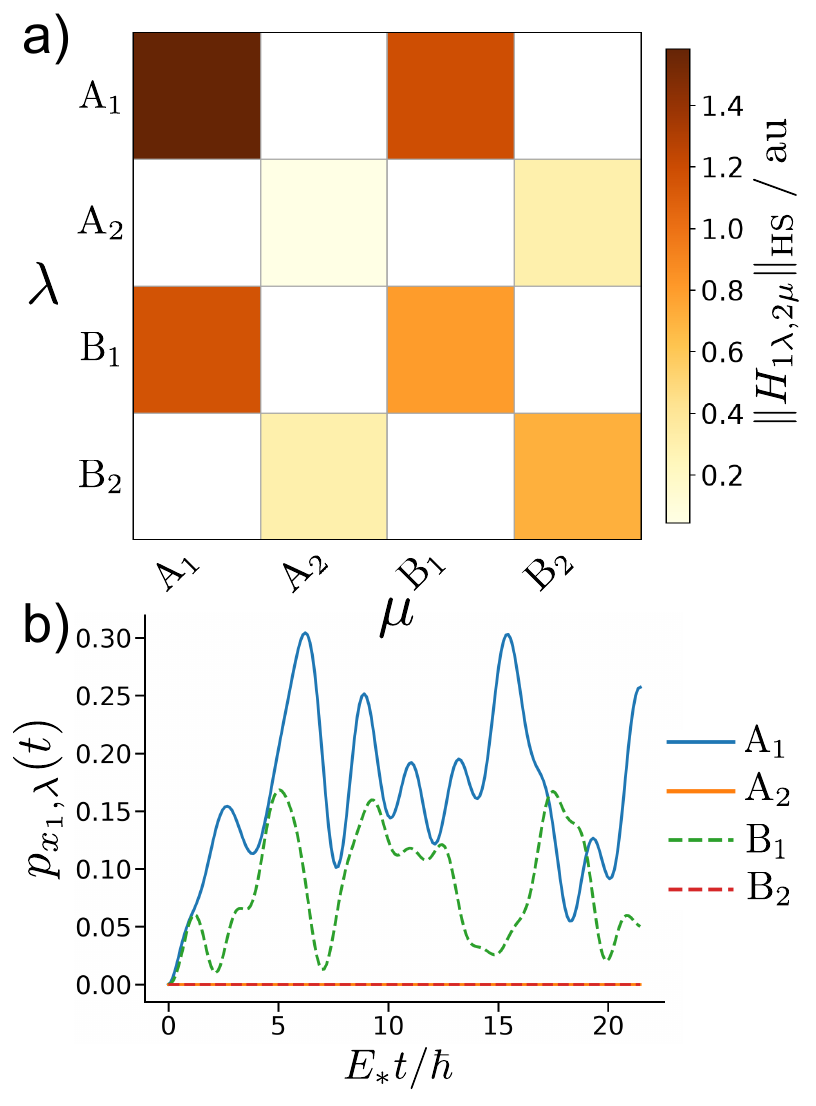}
\caption{Local coupling channels and dynamics in the square aggregate geometry of Figure~\ref{fig:framed-point-groups}. (a) Hilbert--Schmidt norms of the Hamiltonian blocks between local isotypic sectors at two sites, with final labels $\lambda$ on rows and initial labels
$\mu$ on columns. These are the total endpoint norms
$\|H_{1\lambda,2\mu}\|_{\mathrm{HS}}$ of Eq.~\eqref{eq:total-local-channel},
in atomic units. The pairs $\{A_1,B_1\}$ and $\{A_2,B_2\}$ have opposite
reflection parity and do not couple. (b) Populations $p_{x_1,\lambda}(t)$ in the local isotypic sectors at site $x_1$ during evolution from the initial $B_1$ sector at site $x_2$. The $A_2$ and $B_2$ populations remain zero while population transfers into the $A_1$ and $B_1$ sectors. Time is expressed in the dimensionless variable $E_*t/\hbar$, where $E_*$ is the spectral energy scale defined in Appendix~\ref{app:energy-scaling}.}
\label{fig:coupling-dynamics}
\end{figure}
The square configuration illustrates how the character criterion of the
preceding section resolves global dynamical constraints into local
site-to-site exclusions. Its global group $T=C_{4h}$ exchanges four
identically labeled sites, each carrying the local structural symmetry group $P_x=C_{2v}$.
As shown in Appendix~\ref{app:T-recipe}
(Eq.~\ref{eq:square-stabilizer-cocycle}), every site has stabilizer
\[
  T_x=\{e,m\} \subset \mathrm{E}(3),
  \qquad m=(0,M_{xz}),
\]
where $m$ is reflection in the square plane. The square frames give
\[
  \kappa_x(e)=I,
  \qquad \kappa_x(m)=M_{xz}.
\]
Writing $\chi_\lambda^{(x)}=\chi_\lambda\circ\kappa_x$,
Frobenius reciprocity reduces the support calculation to two character
products:
\begin{align*}
  \frac{m_{x,\lambda}^{\gamma}}{n_{x,\lambda}}
  &=\frac12\left[
    \overline{\chi_\gamma(e)}\,\chi_\lambda^{(x)}(e)
    +\overline{\chi_\gamma(m)}\,\chi_\lambda^{(x)}(m)
  \right]\\
  &=\frac12\left[
    1+\overline{\chi_\gamma(m)}\,\chi_\lambda^{(x)}(m)
  \right].
\end{align*}
Here $n_{x,\lambda}>0$ is the multiplicity of the local irrep in the
model sector. The identity contribution
$\overline{\chi_\gamma(e)}\chi_\lambda^{(x)}(e)=1$
because the complex irreducible representations of both groups are
one-dimensional.

For the numerical realization, each site basis consists of spherical
harmonics truncated at $l_{\max}=2$. All four local irreps occur.
Their dimensions and characters pulled back to the aggregate site stabilizer are
\[
\begin{array}{c|c|rr}
  \text{Local irrep} & \dim\widetilde K_{x,\lambda} & e & m\\ \hline
  A_1 & 4 & 1 &  1\\
  A_2 & 1 & 1 & -1\\
  B_1 & 2 & 1 &  1\\
  B_2 & 2 & 1 & -1
\end{array}
\]
The global characters restricted to the same stabilizer are listed below,
together with the dimensions of the global isotypic components in the
36-dimensional truncated aggregate space:
\[
\begin{array}{c|c|rr}
  \text{Global irreps} & \dim W_\gamma & e & m\\ \hline
  A_g,\ B_g,\ E_u^{(+i)},\ E_u^{(-i)} & 6 & 1 &  1\\
  A_u,\ B_u,\ E_g^{(+i)},\ E_g^{(-i)} & 3 & 1 & -1
\end{array}
\]
Each complex irrep has dimension $d_\gamma=1$, so these block dimensions
are also its multiplicities in $W$.
The scalar product therefore equals one when the reflection characters
agree and zero when they differ. The local sectors $A_1,B_1$ have support
only in $A_g,B_g,E_u^{(+i)},E_u^{(-i)}$, whereas $A_2,B_2$ have support
only in $A_u,B_u,E_g^{(+i)},E_g^{(-i)}$. These support sets are disjoint.
Consequently, for endpoints belonging to opposite reflection classes,
\[
  m_{x_i,\lambda}^{\gamma}m_{x_j,\mu}^{\gamma}=0
  \qquad\text{for every }\gamma\in\Irr(T).
\]
The corresponding site-to-site channels vanish under any globally
invariant Hamiltonian. This exclusion is independent of the numerical
couplings and of the local multiplicities, provided the sectors are present.

Figure~\ref{fig:coupling-dynamics}a shows the Hilbert--Schmidt norms of the local
endpoint blocks of the symmetrized Hamiltonian introduced in
Section~\ref{sec:operator-symmetrization} and
Figure~\ref{fig:square-hamiltonian}. The channels connecting opposite
reflection sectors vanish as predicted, while the remaining coupling
strengths depend on the chosen Hamiltonian. The construction of the
adapted bases and reduced operator blocks is given in
Appendix~\ref{app:adapted-bases}.

To compare this exclusion with dynamics, we prepared a normalized state
$\psi_0$ in the $B_1$ local isotypic component at site $x_2$, as visualized in
Figure~\ref{fig:square-orbitals}a. Then, $\psi_0$ was evolved under the aggregate
Hamiltonian detailed in Section~\ref{sec:operator-symmetrization}, while
monitoring symmetry sector populations. The population
of the local $\lambda$-isotypic component on site $x_i$ is given by
\[
p_{x_i,\lambda}(t)
:=\braket{\psi(t)|\overline{\mathcal Q}_{x_i,\lambda}|\psi(t)},
\qquad
\psi(t)=e^{-itH/\hbar}\psi_0.
\]
The sum of these populations over $\lambda$ gives the site population and the
total population---upon summing over all local sectors at all sites---is conserved.
Figure~\ref{fig:coupling-dynamics}b shows the local symmetry sector populations at the target site $x_1$ as a function of dimensionless time $E_*t/\hbar$. The opposite-reflection sectors $A_2$ and $B_2$ remain unpopulated because they have no global support in common with the initial $B_1$ sector, and the propagator preserves the global isotypic components.
\section{Discussion}
\label{sec:discussion}
\proseparagraphs
The question motivating this work was under what conditions symmetry attributed
to a local molecular subunit remains meaningful once that subunit is embedded
within an aggregate. The construction developed here gives a conditional answer.
A local symmetry label need not remain meaningful simply because the
corresponding structural motif can still be identified. Rather, its symmetry
content must admit a covariant realization within the geometry of the complete
system. The relative positions and frames of the local structural symmetry groups determine
the maximal compatible global symmetry $T$, while the transport
cocycle determines how representations attached to different sites are related.
The resulting local-to-global representation therefore places the local and
global symmetry descriptions within a single representation-theoretic
structure.

This distinction clarifies the role of decomposition in models of molecular
aggregates. A site decomposition identifies local degrees of freedom and makes
questions of localization, transfer, and intersite dynamics naturally
expressible, but the decomposition alone does not determine how the symmetry
labels assigned to those sites are related. Here this additional structure is
supplied by the framed geometry and its associated representation. Moreover,
when the ambient group $G$ acts on a physical Hilbert space
$(U,\mathcal H)$, the reconstruction map $F_s$ intertwines
$\rho_{\mathrm{glob}}$ with $U|_{T}$. The local-to-global
representation is therefore not an independent replacement for the global
quantum description: it provides a local resolution whose symmetry content is
compatible with that description.

The purpose is not to assume that every local symmetry distinction survives
within the aggregate, but to determine what inferential content it retains.
Distinct local labels may have identical global support and thus remain
indistinguishable by the exclusion criterion developed here. Conversely,
disjoint support establishes exclusions between specified local preparation
and observation sectors. The refinement therefore concerns what can be
inferred about local endpoints from their compatibility with aggregate symmetry.

This also separates two senses in which local symmetry may be informative.
The first is kinematic. The overlaps
\[
b^\gamma_{x,\lambda}
=
\|\mathcal{P}_\gamma\overline{\mathcal{Q}}_{x,\lambda}\|_{\mathrm{HS}}^2
\]
determine which global symmetry sectors can contain a state localized in the
local sector $(x,\lambda)$. In this sense, a local label survives passage to
the aggregate when it has well-defined support in the common global
representation. Such support, however, is not itself a dynamical selection
rule. Distinct local sectors may occupy the same global isotypic component,
and global symmetry alone cannot distinguish the corresponding multiplicity
degrees of freedom.

A conventional global symmetry analysis establishes that inequivalent global
isotypic components cannot exchange amplitude,
\[
  \mathcal P_\gamma U(t)\mathcal P_{\gamma'}=0
  \qquad(\gamma\neq\gamma').
\]
Here \(U(t)=e^{-itH/\hbar}\) denotes Hamiltonian evolution. This statement
alone does not identify which local preparation and observation sectors
these exclusions separate. Our construction determines the global support
of each local structural symmetry sector, thereby resolving the global
exclusions into site-to-site statements without introducing additional
conserved quantities.

For globally invariant dynamics, this support gives exclusions at the level
of local endpoints: sectors with no shared global support cannot exchange
amplitude. When global symmetries exchange sites, the local sectors remain
invariant under their aggregate site stabilizers but need not be invariant under the
complete group. Induction and Frobenius reciprocity make their global support
accessible through stabilizer character scalar products. The resulting
exclusions express global symmetry in a local resolution and require the
geometric relation between the local and global actions, beyond their abstract
character tables.

Although evaluating this exclusion criterion requires only character data
and the framed stabilizer map $\kappa_x$, its interpretation as a
local-to-global exclusion follows from the induced-representation
construction. Frobenius reciprocity converts the multiplicity in that
induced representation into a scalar product on $T_x$. For a local sector
present in the model, its multiplicity does not affect whether this scalar
product vanishes. The exclusion can therefore be determined without
explicitly constructing either the model representation or its induction.
The Hilbert--Schmidt norms of the resolved Hamiltonian blocks then quantify the
couplings for a particular model.

The refinement established here concerns the local resolution of global
dynamical exclusions. It does not establish additional dynamical constraints
arising from local symmetry operations that are not aggregate symmetries.
Conversely, it does not rule out additional exclusions when further local
structure or assumptions on the Hamiltonian are supplied. Global-support
exclusions therefore need not exhaust the consequences of local structure.

The distinction between physical geometry and fiber-coordinate identifications is
essential to this interpretation. Relative site positions and orientations are
physical data, changing them may alter $T$, the cocycle, the global
support of local sectors, and the resulting Hamiltonian channels.
By contrast, changing the fiber-coordinate identifications with the geometric
datum fixed produces an equivalent coordinate realization of
$\rho_{\mathrm{glob}}$. The
resulting symmetry statements therefore depend on the relative geometry of the
aggregate without depending on arbitrary choices of local coordinates.

The relevance to chirality-induced spin selectivity raised in the Introduction
concerns how local molecular geometry enters the symmetry description of a
complete transport system. Dednam and collaborators demonstrated that
spin-polarization constraints depend on the symmetries of the entire junction,
including the electrodes and their relative orientations.\cite{dednam23_group_theor_approac_to_origin}
The present framework provides a general construction relating local
structural symmetries to the compatible global action. Applying this
construction to spin-dependent transport would additionally require a
dynamical model and a specified spin-resolved transport observable, allowing
the consequences of the framed geometry to be assessed for a particular
measurement.

This framework is complementary to full aggregate calculations. Such
calculations may determine energies, wavefunctions, transition amplitudes, or
time-dependent observables for a particular molecular structure. The present
construction instead supplies a representation-theoretic language in which
relations among local structural units can be posed independently of any one
numerical realization. It thereby formalizes a common chemical mode of
reasoning---assigning states and processes to local structural motifs---while
making explicit the conditions under which that reasoning is compatible with
the symmetry of the aggregate.

\section{Conclusion}
\label{sec:conclusion}
\proseparagraphs
Local symmetry remains meaningful within an aggregate not merely when a
structural subunit can be identified, but when its symmetry data can be
carried covariantly into a representation of the complete system. We have
constructed such a local-to-global representation from framed local
structural symmetry groups, their relative geometry, and a representation of the local
transformation group. The same construction determines the maximal compatible
global symmetry, relates site-resolved local sectors to global isotypic
components, and, through the reconstruction intertwiner, connects this local
description to the physical quantum representation.

For aggregates admitting site exchanges, the construction relates
stabilizer-invariant local sectors to the global isotypic decomposition.
Character scalar products, simplified by Frobenius reciprocity, determine
their global support and hence local dynamical exclusions. Hilbert--Schmidt norms
of the projected Hamiltonian blocks quantify the remaining couplings for a
specified model. This gives a concrete meaning to local symmetry within the
aggregate: it identifies which local endpoints can participate in globally
compatible dynamics while retaining the dependence on geometry and on the
Hamiltonian itself.

\section*{Acknowledgements}
This work was supported as part of the Center for Molecular Quantum Transduction,
an Energy Frontier Research Center funded by the U.S. Department of Energy,
Office of Science, Basic Energy Sciences, under Award \#DE-SC0021314.

\printbibliography

@book{primas83_chemis_quant_mechan_reduc,
  author    = {Hans Primas},
  title     = {Chemistry, Quantum Mechanics and Reductionism},
  year      = {1983},
  publisher = {Springer Berlin Heidelberg},
  doi       = {10.1007/978-3-642-69365-6}
}

@article{katritzky10_quant_correl_physic_chemic_proper,
  author  = {Alan R. Katritzky and Minati Kuanar and Svetoslav Slavov and C. Dennis Hall and Mati Karelson and Iiris Kahn and Dimitar A. Dobchev},
  title   = {Quantitative Correlation of Physical and Chemical Properties with Chemical Structure: Utility for Prediction},
  journal = {Chemical Reviews},
  volume  = {110},
  number  = {10},
  pages   = {5714--5789},
  year    = {2010},
  doi     = {10.1021/cr900238d}
}

@article{hendry23_struc_essen_exist_chemis,
  author  = {Robin Findlay Hendry},
  title   = {Structure, Essence and Existence in Chemistry},
  journal = {Ratio},
  volume  = {36},
  number  = {4},
  pages   = {274--288},
  year    = {2023},
  doi     = {10.1111/rati.12387}
}

@article{talanquer18_progr_reason_about_struc_proper_relat,
  author  = {Vicente Talanquer},
  title   = {Progressions in Reasoning About Structure--Property Relationships},
  journal = {Chemistry Education Research and Practice},
  volume  = {19},
  number  = {4},
  pages   = {998--1009},
  year    = {2018},
  doi     = {10.1039/C7RP00187H}
}

@article{musil21_physic_inspir_struc_repres_molec_mater,
  author  = {Felix Musil and Andrea Grisafi and Albert P. Bart{\'o}k and Christoph Ortner and G{\'a}bor Cs{\'a}nyi and Michele Ceriotti},
  title   = {Physics-Inspired Structural Representations for Molecules and Materials},
  journal = {Chemical Reviews},
  volume  = {121},
  number  = {16},
  pages   = {9759--9815},
  year    = {2021},
  doi     = {10.1021/acs.chemrev.1c00021}
}

@book{nitzan2013chemical,
  title={Chemical Dynamics in Condensed Phases: Relaxation, Transfer, and Reactions in Condensed Molecular Systems},
  author={Nitzan, A.},
  isbn={9780199686681},
  lccn={2005030160},
  series={Oxford Graduate Texts},
  url={https://books.google.com/books?id=jrmznAEACAAJ},
  year={2013},
  publisher={OUP Oxford}
}

@article{prodan05_nears_elect_matter,
  author =	 {E. Prodan and W. Kohn},
  title =	 {Nearsightedness of Electronic Matter},
  journal =	 {Proceedings of the National Academy of Sciences},
  volume =	 102,
  number =	 33,
  pages =	 {11635-11638},
  year =	 2005,
  doi =		 {10.1073/pnas.0505436102},
  url =		 {http://dx.doi.org/10.1073/pnas.0505436102},
  DATE_ADDED =	 {Thu Oct 22 15:40:44 2026},
}

@article{fias17_chemic_trans_funct_group_follow,
  author =	 {Stijn Fias and Farnaz Heidar-Zadeh and Paul
                  Geerlings and Paul W. Ayers},
  title =	 {Chemical Transferability of Functional Groups
                  Follows From the Nearsightedness of Electronic
                  Matter},
  journal =	 {Proceedings of the National Academy of Sciences},
  volume =	 114,
  number =	 44,
  pages =	 {11633-11638},
  year =	 2017,
  doi =		 {10.1073/pnas.1615053114},
  url =		 {http://dx.doi.org/10.1073/pnas.1615053114},
  DATE_ADDED =	 {Thu Oct 22 16:22:48 2026},
}

@article{raghavan-chitra26_permut_symmet_molec_aggreg,
  author =	 {Sricharan Raghavan-Chitra and Arghadip Koner and
                  Joel Yuen-Zhou},
  title =	 {Permutationally Symmetric Molecular Aggregates},
  journal =	 {The Journal of Chemical Physics},
  volume =	 165,
  number =	 8,
  pages =	 084306,
  year =	 2026,
  doi =		 {10.1063/5.0334324},
  url =		 {http://dx.doi.org/10.1063/5.0334324},
  DATE_ADDED =	 {Thu Oct 22 16:38:59 2026},
}

@article{bradlyn17_topol_quant_chemis,
  author =	 {Barry Bradlyn and L. Elcoro and Jennifer Cano and
                  M. G. Vergniory and Zhijun Wang and C. Felser and
                  M. I. Aroyo and B. Andrei Bernevig},
  title =	 {Topological Quantum Chemistry},
  journal =	 {Nature},
  volume =	 547,
  number =	 7663,
  pages =	 {298-305},
  year =	 2017,
  doi =		 {10.1038/nature23268},
  url =		 {http://dx.doi.org/10.1038/nature23268},
  DATE_ADDED =	 {Thu Oct 22 23:19:33 2026},
}

@article{evarestov84_applic_band_repres_space_group,
  author =	 {R. A. Evarestov and V. P. Smirnov},
  title =	 {Application of the Band Representations of Space
                  Groups in the Theory of Electronic States of
                  Crystalline Solids. I. General Consideration of the
                  Band Representations},
  journal =	 {physica status solidi (b)},
  volume =	 122,
  number =	 1,
  pages =	 {231-238},
  year =	 1984,
  doi =		 {10.1002/pssb.2221220127},
  url =		 {http://dx.doi.org/10.1002/pssb.2221220127},
  DATE_ADDED =	 {Fri Oct 23 15:26:07 2026},
}

@book{evarestov97_site_symmet_cryst,
  author =	 {Robert A. Evarestov and Vyacheslav P. Smirnov},
  title =	 {Site Symmetry in Crystals},
  year =	 1997,
  publisher =	 {Springer Berlin Heidelberg},
  url =		 {http://dx.doi.org/10.1007/978-3-642-60488-1},
  DATE_ADDED =	 {Fri Oct 23 12:59:54 2026},
  doi =		 {10.1007/978-3-642-60488-1},
  pages =	 {nil},
  series =	 {Springer Series in Solid-State Sciences},
}

@article{cano21_band_repres_topol_quant_chemis,
  author =	 {Jennifer Cano and Barry Bradlyn},
  title =	 {Band Representations and Topological Quantum
                  Chemistry},
  journal =	 {Annual Review of Condensed Matter Physics},
  volume =	 12,
  number =	 1,
  pages =	 {225-246},
  year =	 2021,
  doi =		 {10.1146/annurev-conmatphys-041720-124134},
  url =
                  {http://dx.doi.org/10.1146/annurev-conmatphys-041720-124134},
  DATE_ADDED =	 {Fri Oct 23 15:58:15 2026},
}

@article{cano18_build_block_topol_quant_chemis,
  author =	 {Jennifer Cano and Barry Bradlyn and Zhijun Wang and
                  L. Elcoro and M. G. Vergniory and C. Felser and
                  M. I. Aroyo and B. Andrei Bernevig},
  title =	 {Building Blocks of Topological Quantum Chemistry:
                  Elementary Band Representations},
  journal =	 {Physical Review B},
  volume =	 97,
  number =	 3,
  pages =	 035139,
  year =	 2018,
  doi =		 {10.1103/physrevb.97.035139},
  url =		 {http://dx.doi.org/10.1103/physrevb.97.035139},
  DATE_ADDED =	 {Fri Oct 23 13:10:08 2026},
}

@incollection{KUHN201147,
title = {Chapter 2 - Quantum Dynamics and Spectroscopy of Excitons in Molecular Aggregates},
editor = {Uli Wüerfel and Michael Thorwart and Eicke R. Weber},
series = {Semiconductors and Semimetals},
publisher = {Elsevier},
volume = {85},
pages = {47-81},
year = {2011},
booktitle = {Quantum Efficiency in Complex Systems, Part II},
issn = {0080-8784},
doi = {https://doi.org/10.1016/B978-0-12-391060-8.00002-2},
url = {https://www.sciencedirect.com/science/article/pii/B9780123910608000022},
author = {Oliver Kühn and Stefan Lochbrunner},
}

@article{hestand18_expan_theor_h_j_molec_aggreg,
  author =	 {Nicholas J. Hestand and Frank C. Spano},
  title =	 {Expanded Theory of H- and J-Molecular Aggregates:
                  the Effects of Vibronic Coupling and Intermolecular
                  Charge Transfer},
  journal =	 {Chemical Reviews},
  volume =	 118,
  number =	 15,
  pages =	 {7069-7163},
  year =	 2018,
  doi =		 {10.1021/acs.chemrev.7b00581},
  url =		 {http://dx.doi.org/10.1021/acs.chemrev.7b00581},
  DATE_ADDED =	 {Fri Oct 23 16:38:29 2026},
}

@book{Berndt2007,
  title     = {Representations of Linear Groups: An Introduction Based on Examples from Physics and Number Theory},
  author    = {Rolf Berndt},
  year      = {2007},
  publisher = {Vieweg+Teubner Verlag Wiesbaden},
  isbn      = {978-3-8348-0319-1},
  doi       = {10.1007/978-3-8348-9401-4},
  address   = {Wiesbaden, Germany},
  edition   = {1}
}

@article{eckart30_applic_group_theor_to_quant,
  author        = {Carl Eckart},
  title         = {The Application of Group Theory To the Quantum Dynamics of Monatomic Systems},
  journal       = {Reviews of Modern Physics},
  volume        = 2,
  number        = 3,
  pages         = {305--380},
  year          = 1930,
  doi           = {10.1103/revmodphys.2.305},
  url           = {http://dx.doi.org/10.1103/RevModPhys.2.305},
  date_added    = {Mon Aug 4 13:53:34 2025}
}

@book{wigner2013group,
  title         = {Group Theory: And Its Application to the Quantum Mechanics of Atomic Spectra},
  author        = {Wigner, E.P. and Massey, H.S.W.},
  isbn          = {9781483275765},
  url           = {https://books.google.com/books?id=UITNCgAAQBAJ},
  year          = {2013},
  publisher     = {Academic Press}
}

@book{woit17_quant_theor_group_repres,
  author =	 {Peter Woit},
  title =	 {Quantum Theory, Groups and Representations},
  year =	 2017,
  publisher =	 {Springer International Publishing},
  url =		 {http://dx.doi.org/10.1007/978-3-319-64612-1},
  DATE_ADDED =	 {Fri Oct 23 17:04:15 2026},
  doi =		 {10.1007/978-3-319-64612-1},
}

@article{garzon-ramirez26_how_symmet_gover_dihed_angle,
  author =	 {Antonio J. Garz{\'o}n-Ram{\'i}rez and Connor
                  K. Terry Weatherly and Kyle T. Kairys and Michael
                  R. Wasielewski and Roel Tempelaar},
  title =	 {How Symmetry Governs the Dihedral Angle Dependence
                  of Intermolecular Spin-Orbit Coupling},
  journal =	 {The Journal of Physical Chemistry Letters},
  volume =	 17,
  number =	 28,
  pages =	 {7887-7892},
  year =	 2026,
  doi =		 {10.1021/acs.jpclett.6c01656},
  url =		 {http://dx.doi.org/10.1021/acs.jpclett.6c01656},
  DATE_ADDED =	 {Fri Oct 23 17:36:07 2026},
}

@article{peeks17_elect_deloc_radic_cation_porph,
  author =	 {Martin D. Peeks and Claudia E. Tait and Patrik
                  Neuhaus and Georg M. Fischer and Markus Hoffmann and
                  Renée Haver and Arjen Cnossen and Jeffrey R. Harmer
                  and Christiane R. Timmel and Harry L. Anderson},
  title =	 {Electronic Delocalization in the Radical Cations of
                  Porphyrin Oligomer Molecular Wires},
  journal =	 {Journal of the American Chemical Society},
  volume =	 139,
  number =	 30,
  pages =	 {10461-10471},
  year =	 2017,
  doi =		 {10.1021/jacs.7b05386},
  url =		 {http://dx.doi.org/10.1021/jacs.7b05386},
  DATE_ADDED =	 {Fri Oct 23 17:47:01 2026},
}

@article{plasser12_analy_excit_charg_trans_inter,
  author =	 {Felix Plasser and Hans Lischka},
  title =	 {Analysis of Excitonic and Charge Transfer
                  Interactions From Quantum Chemical Calculations},
  journal =	 {Journal of Chemical Theory and Computation},
  volume =	 8,
  number =	 8,
  pages =	 {2777-2789},
  year =	 2012,
  doi =		 {10.1021/ct300307c},
  url =		 {http://dx.doi.org/10.1021/ct300307c},
  DATE_ADDED =	 {Fri Oct 23 18:08:55 2026},
}

@article{nitzan03_elect_trans_molec_wire_junct,
  author =	 {Abraham Nitzan and Mark A. Ratner},
  title =	 {Electron Transport in Molecular Wire Junctions},
  journal =	 {Science},
  volume =	 300,
  number =	 5624,
  pages =	 {1384-1389},
  year =	 2003,
  doi =		 {10.1126/science.1081572},
  url =		 {http://dx.doi.org/10.1126/science.1081572},
  DATE_ADDED =	 {Fri Oct 23 18:12:08 2026},
}

@article{dagotto94_correl_elect_high_temper_super,
  author =	 {Elbio Dagotto},
  title =	 {Correlated Electrons in High-Temperature
                  Superconductors},
  journal =	 {Reviews of Modern Physics},
  volume =	 66,
  number =	 3,
  pages =	 {763-840},
  year =	 1994,
  doi =		 {10.1103/revmodphys.66.763},
  url =		 {http://dx.doi.org/10.1103/revmodphys.66.763},
  DATE_ADDED =	 {Fri Oct 23 18:13:29 2026},
}

@article{guo12_spin_selec_trans_elect_dna_doubl_helix,
  author =	 {Ai-Min Guo and Qing-feng Sun},
  title =	 {Spin-Selective Transport of Electrons in Dna Double
                  Helix},
  journal =	 {Physical Review Letters},
  volume =	 108,
  number =	 21,
  pages =	 218102,
  year =	 2012,
  doi =		 {10.1103/physrevlett.108.218102},
  url =		 {http://dx.doi.org/10.1103/physrevlett.108.218102},
  DATE_ADDED =	 {Fri Oct 23 18:18:00 2026},
}

@article{spano03_fundam_photop_conjug_oligom_herrin_aggreg,
  author =	 {Frank C. Spano},
  title =	 {The Fundamental Photophysics of Conjugated Oligomer
                  Herringbone Aggregates},
  journal =	 {The Journal of Chemical Physics},
  volume =	 118,
  number =	 2,
  pages =	 {981-994},
  year =	 2003,
  doi =		 {10.1063/1.1521933},
  url =		 {http://dx.doi.org/10.1063/1.1521933},
  DATE_ADDED =	 {Fri Oct 23 18:31:22 2026},
}

@article{spano10_vibron_coupl_j_aggreg_beyon,
  author =	 {Frank C. Spano and Hajime Yamagata},
  title =	 {Vibronic Coupling in J-Aggregates and Beyond: a
                  Direct Means of Determining the Exciton Coherence
                  Length From the Photoluminescence Spectrum},
  journal =	 {The Journal of Physical Chemistry B},
  volume =	 115,
  number =	 18,
  pages =	 {5133-5143},
  year =	 2010,
  doi =		 {10.1021/jp104752k},
  url =		 {http://dx.doi.org/10.1021/jp104752k},
  DATE_ADDED =	 {Fri Oct 23 18:32:50 2026},
}

@article{spano91_excit_confin_geomet,
  author =	 {Francis C. Spano and Shaul Mukamel},
  title =	 {Excitons in Confined Geometries: Size Scaling of
                  Nonlinear Susceptibilities},
  journal =	 {The Journal of Chemical Physics},
  volume =	 95,
  number =	 10,
  pages =	 {7526-7540},
  year =	 1991,
  doi =		 {10.1063/1.461379},
  url =		 {http://dx.doi.org/10.1063/1.461379},
  DATE_ADDED =	 {Fri Oct 23 18:35:56 2026},
}

@article{evers22_theor_chiral_induc_spin_selec,
  author =	 {Ferdinand Evers and Amnon Aharony and Nir Bar‐Gill
                  and Ora Entin‐Wohlman and Per Hedeg{\aa}rd and Oded
                  Hod and Pavel Jelinek and Grzegorz Kamieniarz and
                  Mikhail Lemeshko and Karen Michaeli and Vladimiro
                  Mujica and Ron Naaman and Yossi Paltiel and Sivan
                  Refaely‐Abramson and Oren Tal and Jos Thijssen and
                  Michael Thoss and Jan M. van Ruitenbeek and Latha
                  Venkataraman and David H. Waldeck and Binghai Yan
                  and Leeor Kronik},
  title =	 {Theory of Chirality Induced Spin Selectivity:
                  Progress and Challenges},
  journal =	 {Advanced Materials},
  volume =	 34,
  number =	 13,
  pages =	 2106629,
  year =	 2022,
  doi =		 {10.1002/adma.202106629},
  url =		 {http://dx.doi.org/10.1002/adma.202106629},
  DATE_ADDED =	 {Fri Oct 23 18:45:49 2026},
}

@book{husemoller94_fibre_bundl,
  author =	 {Dale Husemoller},
  title =	 {Fibre Bundles},
  year =	 1994,
  publisher =	 {Springer New York},
  url =		 {http://dx.doi.org/10.1007/978-1-4757-2261-1},
  DATE_ADDED =	 {Fri Oct 23 19:25:16 2026},
  doi =		 {10.1007/978-1-4757-2261-1},
  pages =	 {nil},
  series =	 {Graduate Texts in Mathematics},
}

@book{serre77_linear_repres_finit_group,
  author =	 {Jean-Pierre Serre},
  title =	 {Linear Representations of Finite Groups},
  year =	 1977,
  publisher =	 {Springer New York},
  url =		 {http://dx.doi.org/10.1007/978-1-4684-9458-7},
  DATE_ADDED =	 {Fri Oct 23 20:58:55 2026},
  doi =		 {10.1007/978-1-4684-9458-7},
  pages =	 {nil},
  series =	 {Graduate Texts in Mathematics},
}

@incollection{STEINBORN19731,
title = {Rotation and Translation of Regular and Irregular Solid Spherical Harmonics**Work was performed in part in the Ames Laboratory of the United States Atomic Energy Commission.},
editor = {Per-Olov Löwdin},
series = {Advances in Quantum Chemistry},
publisher = {Academic Press},
volume = {7},
pages = {1-81},
year = {1973},
issn = {0065-3276},
doi = {https://doi.org/10.1016/S0065-3276(08)60558-4},
url = {https://www.sciencedirect.com/science/article/pii/S0065327608605584},
author = {E.O. Steinborn and K. Ruedenberg},
}

@book{Cotton1990-td,
  title     = {Chemical Applications of Group Theory},
  author    = {Cotton, F. Albert},
  publisher = {John Wiley \& Sons},
  edition   = {3},
  year      = {1990},
  location  = {New York},
  isbn      = {9780471510949},
  langid    = {english}
}

@book{derksen15_comput_invar_theor,
  author =	 {Harm Derksen and Gregor Kemper},
  title =	 {Computational Invariant Theory},
  year =	 2015,
  publisher =	 {Springer Berlin Heidelberg},
  url =		 {http://dx.doi.org/10.1007/978-3-662-48422-7},
  DATE_ADDED =	 {Fri Oct 23 22:36:22 2026},
  doi =		 {10.1007/978-3-662-48422-7},
  pages =	 {nil},
  series =	 {Encyclopaedia of Mathematical Sciences},
}

@article{dednam23_group_theor_approac_to_origin,
  author =	 {W. Dednam and M. A. Garc{\'i}a-Bl{\'a}zquez and
                  Linda A. Zotti and E. B. Lombardi and C. Sabater and
                  S. Pakdel and J. J. Palacios},
  title =	 {A Group-Theoretic Approach To the Origin of
                  Chirality-Induced Spin-Selectivity in Nonmagnetic
                  Molecular Junctions},
  journal =	 {ACS Nano},
  volume =	 17,
  number =	 7,
  pages =	 {6452-6465},
  year =	 2023,
  doi =		 {10.1021/acsnano.2c11410},
  url =		 {http://dx.doi.org/10.1021/acsnano.2c11410},
}

@article{littlejohn97_gauge_field_separ_rotat_andin,
  author =	 {Robert G. Littlejohn and Matthias Reinsch},
  title =	 {Gauge Fields in the Separation of Rotations
                  Andinternal Motions in the N-Body Problem},
  journal =	 {Reviews of Modern Physics},
  volume =	 69,
  number =	 1,
  pages =	 {213-276},
  year =	 1997,
  doi =		 {10.1103/revmodphys.69.213},
  url =		 {http://dx.doi.org/10.1103/revmodphys.69.213},
  DATE_ADDED =	 {Tue Nov 10 14:57:24 2026},
}

\appendix
\mathematicalparagraphs

\section{Group property and maximality of \texorpdfstring{$T$}{T}}
\label{app:group}
Let
\[
  S:=\left\{g\in T_0\,\middle|\,
  \kappa(g,x)P_x\kappa(g,x)^{-1}=P_{\sigma_g(x)}
  \text{ for all }x\in X\right\}.
\]
This is the set on the right-hand side of
Eq.~\eqref{eq:global-group} without the condition
$\kappa(g,x)\in P_{\sigma_g(x)}$. We first show that $S$ is a subgroup
of $T_0$.

\paragraph{Identity.}
Since $\kappa(e,x)=e$ and $\sigma_e(x)=x$ for every $x\in X$, the
identity lies in $S$.

\paragraph{Closure.}
If $g_1,g_2\in S$, the cocycle identity
(Eq.~\eqref{eq:cocycle-identity}) gives
\begin{align*}
 &\kappa(g_1g_2,x)P_x\kappa(g_1g_2,x)^{-1}\\
 &\quad=
 \kappa(g_1,\sigma_{g_2}(x))
 \bigl(\kappa(g_2,x)P_x\kappa(g_2,x)^{-1}\bigr)
 \kappa(g_1,\sigma_{g_2}(x))^{-1}\\
 &\quad=P_{\sigma_{g_1}(\sigma_{g_2}(x))}
 =P_{\sigma_{g_1g_2}(x)}.
\end{align*}
Thus $g_1g_2\in S$.

\paragraph{Inverses.}
For $g\in S$, its defining condition implies
\[
  \kappa(g,x)^{-1}P_{\sigma_g(x)}\kappa(g,x)=P_x.
\]
Setting $g_1=g^{-1}$ and $g_2=g$ in the cocycle identity gives
\[
  \kappa(g^{-1},\sigma_g(x))=\kappa(g,x)^{-1}.
\]
Consequently,
\[
  \kappa(g^{-1},\sigma_g(x))P_{\sigma_g(x)}
  \kappa(g^{-1},\sigma_g(x))^{-1}
  =P_x=P_{\sigma_{g^{-1}}(\sigma_g(x))}.
\]
Since $\sigma_g$ permutes $X$, this establishes the defining condition
of $S$ for $g^{-1}$ at every site. Hence $g^{-1}\in S$, and $S\leq T_0$.

\paragraph{Preservation of the embedded groups.}
Using $gs(x)=s(\sigma_g(x))\kappa(g,x)$, we obtain
\[
  g\widetilde P_xg^{-1}
  =s(\sigma_g(x))
    \bigl(\kappa(g,x)P_x\kappa(g,x)^{-1}\bigr)
    s(\sigma_g(x))^{-1}.
\]
Therefore $g\in T_0$ belongs to $S$ if and only if
$g\widetilde P_xg^{-1}=\widetilde P_{\sigma_g(x)}$ for every $x\in X$.
Thus $S$ is precisely the maximal subgroup of $T_0$ preserving the
embedded groups $\{\widetilde P_x\}_{x\in X}$ under conjugation.

\paragraph{Preservation of the frame cosets.}
Write $C_x:=s(x)P_x$. For $g\in S$,
\[
  gC_x=s(\sigma_g(x))\kappa(g,x)P_x
      =s(\sigma_g(x))P_{\sigma_g(x)}\kappa(g,x),
\]
where the second equality uses the defining condition of $S$. Hence
\[
  gC_x=C_{\sigma_g(x)}
  \quad\Longleftrightarrow\quad
  \kappa(g,x)\in P_{\sigma_g(x)}.
\]
This is precisely the additional condition defining $T\subseteq S$.
The identity preserves every $C_x$. If $g_1,g_2\in S$ preserve the frame
cosets at every site, then
\[
  (g_1g_2)C_x
  =g_1C_{\sigma_{g_2}(x)}
  =C_{\sigma_{g_1}(\sigma_{g_2}(x))}
  =C_{\sigma_{g_1g_2}(x)}.
\]
If $g\in S$ preserves the frame cosets at every site, then
\[
  gC_{\sigma_g^{-1}(x)}=C_x
  \quad\Longrightarrow\quad
  g^{-1}C_x=C_{\sigma_g^{-1}(x)}
  =C_{\sigma_{g^{-1}}(x)}.
\]
Thus these elements form a subgroup $T\leq S$, characterized by
Eq.~\eqref{eq:global-group}.

\paragraph{Maximality.}
Preservation of the frame cosets also implies preservation of the embedded
local groups, since
\[
  \widetilde P_x=C_xC_x^{-1}.
\]
Every subgroup of $T_0$ preserving the framed local-group data is therefore
contained in $S$ and, by the preceding equivalence, in $T$. Conversely,
every element of $T$ preserves these data. This proves the maximality
asserted in Theorem~\ref{thm:glob-sym}.

\section{Euclidean recipe for finite local structural symmetry groups}
\label{app:T-recipe}

\paragraph{Step 1) -- Setup.}
Specify the system's geometry by writing the distinct site positions
as $X=\{b_1,\ldots,b_N\}\subset\R^3$ and their frames as
$s_i=(b_i,A_i)\in\R^3\rtimes \mathrm{O}(3)$, and associate to each site a
finite matrix group $P_i\leq \mathrm{O}(3)$. Our convention for the
$\mathrm{E}(3)$ action on $\R^3$ is $(b,A)x=Ax+b$.

Assign each site a label $l:X\to L$, where sites sharing the same
label are prescribed to have identical reference geometric objects.
The labels encode geometric equivalence as input data. In particular,
distinct geometric objects may possess the same point group.

For convenience, sites sharing a label are assigned the same
reference geometric object in a common reference orientation, so
their local structural symmetry groups are equal as matrix subgroups of $\mathrm{O}(3)$:
\[
  l(b_i)=l(b_j)\implies P_i=P_j.
\]
Here equality means equality of the sets of matrices representing
the groups in the common reference coordinates, rather than
merely isomorphic or conjugate point groups.

\paragraph{Step 2) -- Construct Candidate Global Transformations.}
The global symmetry group $T$ may exchange sites and therefore
cannot be obtained solely from the lifted groups
$\widetilde{P}_{b_i}=s_iP_i s_i^{-1}$. Instead, fix one source site $i$
and enumerate its possible targets $j$ satisfying $l(b_j)=l(b_i)$.
The image of this one site and its local transport element determine
the entire Euclidean transformation.

Indeed, an admissible $g=(t,R)$ mapping $b_i$ to $b_j$ satisfies
$Rb_i+t=b_j$ and, by the definition of $T$ in Eq.~\ref{eq:global-group},
\[
  \kappa(g,b_i)=A_j^{-1}RA_i\in P_j.
\]
For each $p\in P_j$, interpreted as a possible value of this cocycle,
solve these two conditions to obtain the orthogonal component
\begin{equation}
\label{eq:orthogonal-component}
  R_{j,p}:=A_jpA_i^{-1},
\end{equation}
and the translational component
\begin{equation}
\label{eq:translational-component}
  t_{j,p}:=b_j-R_{j,p}b_i.
\end{equation}
Together, these define the candidate Euclidean transformation
\begin{equation}
\label{eq:euclidean-candidates}
  g_{j,p}:=(t_{j,p},R_{j,p}).
\end{equation}
Every candidate maps the chosen source to its target with transport
cocycle $p$. No permutations are enumerated in advance: the action on
the remaining sites is determined by the candidate and checked next.

\paragraph{Step 3) -- Filter Candidates.}
Compatibility at the chosen source site does not guarantee compatibility
at the other sites. Apply $g_{j,p}$ to every position and reject it if any
image is absent from $X$. Since the positions are distinct and the candidate
is invertible, a candidate mapping $X$ into itself induces a unique
permutation $\sigma_{j,p}$. Require, for every site $k$,
\begin{align}
\label{eq:euclidean-position-test}
  R_{j,p}b_k+t_{j,p}&=b_{\sigma_{j,p}(k)},\\
\label{eq:euclidean-label-test}
  l(b_{\sigma_{j,p}(k)})&=l(b_k),\\
\label{eq:euclidean-local-test}
  \kappa(g_{j,p},b_k):=A_{\sigma_{j,p}(k)}^{-1}R_{j,p}A_k
    &\in P_{\sigma_{j,p}(k)},\\
\label{eq:euclidean-conjugation-test}
  \kappa(g_{j,p},b_k)P_k\kappa(g_{j,p},b_k)^{-1}
    &=P_{\sigma_{j,p}(k)}.
\end{align}
Thus the candidate must preserve the positions, the site types, and the
local transport conditions at every site. Remove duplicates from the
retained transformations to obtain $T$.

Since label-preserving permutations exchange sites with identical
reference objects, $P_k=P_{\sigma_{j,p}(k)}$. Consequently, the local
transport condition \eqref{eq:euclidean-local-test} automatically
implies the conjugation condition \eqref{eq:euclidean-conjugation-test}.
We retain the latter explicitly to exhibit the defining conditions
of $T$.

\paragraph{Completeness.}
Let $g=(t,R)\in T$ be any admissible transformation and let
$b_j=gb_i$ be the image of the fixed source site. Label preservation
ensures that $j$ is one of the enumerated targets. Its transport cocycle
\[
  p:=\kappa(g,b_i)=A_j^{-1}RA_i\in P_j
\]
uniquely determines $R=A_jpA_i^{-1}=R_{j,p}$. The position condition
then gives $t=b_j-Rb_i=t_{j,p}$. Hence $g=g_{j,p}$, so every admissible
transformation appears among the candidates in Step 2. Since Step 3
tests precisely the defining conditions together with label preservation,
the retained transformations constitute the full global symmetry group
of the labeled configuration. There are $\sum_{j:\,l(b_j)=l(b_i)}|P_j|$ candidates before filtering.

\paragraph{Global point group.}
Every retained transformation fixes the centroid $c=N^{-1}\sum_j b_j$.
Consequently $t=c-Rc$, and $g\mapsto R$ identifies $T$ with a
finite subgroup of $\mathrm{O}(3)$ acting about $c$.

\paragraph{Four-site square.}
The square geometry of Figure~\ref{fig:framed-point-groups}
can be constructed from successive powers of a single generator
\[
  g_0=\left(r\, e_z,R_y(\pi/2)\right)\in \mathrm{E}(3), \qquad r > 0.
\]
For $n=1,2,3,4$, the site frames are
\[
  s(x_n)=g_0^{n-1}=(b_n,A_n)
\]
with
\begin{equation}
\label{eq:square-recipe-data}
  b_n=r\sum_{k=0}^{n-2}R_y(k\pi/2)e_z,
  \qquad
  A_n=R_y((n-1)\pi/2).
\end{equation}
The empty sum for $n=1$ is zero.
These frames describe four identically labeled sites
arranged at the vertices of a square in the $xz$-plane. Each site is given the same label and associated with each site
is the reference local structural symmetry group
\[
P=\{I,C_{2z},M_{xz},M_{yz}\}\cong C_{2v},
\]
where
\[
\begin{aligned}
C_{2z}&=\operatorname{diag}(-1,-1,1),\\
M_{xz}&=\operatorname{diag}(1,-1,1),\\
M_{yz}&=\operatorname{diag}(-1,1,1).
\end{aligned}
\]

To construct the global symmetry group $T$, we apply
the candidate-generation procedure with $x_1$ as the fixed
source site. Since $s(x_1)=e$, each candidate is uniquely
determined by a target site $x_j$ and a reference local structural symmetry-group
element $p\in P$, giving
\[
  g_{j,p}=(b_j,A_jp)=g_0^{j-1}(0,p),
  \qquad j=1,2,3,4.
\]
Thus, there are $4|P|=16$ candidates before filtering.

Since \(g_0\) cyclically permutes the sites (with $x_5=x_1$):
\[
g_0\cdot x_n=x_{n+1},
\]
its transport cocycle is trivial:
\[
\begin{aligned}
\kappa(g_0,x_n)
&=s(x_{n+1})^{-1}g_0\,s(x_n)\\
&=(g_0^n)^{-1}g_0\,g_0^{n-1}\\
&=I_3.
\end{aligned}
\]
Hence $g_0\in T$, and the admissibility of $g_{j,p}$
reduces to that of $(0,p)$.

Of the four elements of $P$, only $I$ and $M_{xz}$
preserve the site positions. Indeed, both fix every
vertex, whereas $M_{yz}$ and $C_{2z}$ send
$b_3=r(e_x+e_z)$ to $r(-e_x+e_z)\notin X$.
Furthermore, since $M_{xz}$ commutes with
$A_n=R_y(n\pi/2)$, its transport cocycle satisfies
\[
  \kappa((0,M_{xz}),x_n)
  =A_n^{-1}M_{xz}A_n=M_{xz}\in P.
\]
Consequently, the filtering procedure retains precisely
the eight transformations
\[
  T
  =\{g_0^j m^\epsilon:
    j=0,1,2,3,\ \epsilon=0,1\},
  \qquad m=(0,M_{xz}).
\]
Since $g_0^4=m^2=e$ and $g_0m=mg_0$, we conclude that
\[
  T=\langle g_0,m\rangle
  \cong C_4\times C_2=C_{4h}.
\]
Only $e$ and $m$ fix a chosen site, so $T_{x_n}=\{e,m\}$ for every $n$.
The stabilizer cocycle values used in the character calculation are
\begin{equation}
\label{eq:square-stabilizer-cocycle}
  \kappa(e,x_n)=I,
  \qquad
  \kappa(m,x_n)=M_{xz}.
\end{equation}

\section{Fiber-coordinate changes and cocycle equivalence}
\label{app:frames}
\begin{proof}[Proof of Theorem~\ref{thm:frame-independence}]
Fix the geometric datum $\mathcal D$ and the associated fibers $K_x$.
Replace the fiber-coordinate identifications $\phi_{s(x)}$ by
\[
  \phi_{s(x)h_x}=\phi_{s(x)}\circ\rho(h_x),
  \qquad h_x\in H.
\]
Every identification induced by a point of $q^{-1}(x)$ has this form,
because $q^{-1}(x)$ is an $H$-torsor. The intrinsic transport maps
$\tau_x(g)$ remain unchanged. Their new coordinate expression is
$\rho(\kappa'(g,x))$, where
\begin{align}
  \kappa'(g,x)
  &:=(s(\sigma_g(x))h_{\sigma_g(x)})^{-1}g\,s(x)h_x\\
  &=h_{\sigma_g(x)}^{-1}\kappa(g,x)h_x.
\label{eq:cocycle-gauge-change}
\end{align}
Therefore the global representation in the new fiber coordinates satisfies
\begin{equation}
  \bigl(\rho'_{\mathrm{glob}}(g)\psi\bigr)_{\sigma_g(x)}
  =\rho(h_{\sigma_g(x)})^{-1}
   \rho(\kappa(g,x))\rho(h_x)\psi_x.
\end{equation}
In the block-coordinate realization of $W$, define the block-diagonal
automorphism $J$ by
\begin{equation}
  (J\psi)_x=\rho(h_x)\psi_x,
  \qquad
  (J^{-1}\psi)_x=\rho(h_x)^{-1}\psi_x.
\end{equation}
Then
\begin{align*}
 \left(J^{-1}\rho_{\mathrm{glob}}(g)J\psi\right)_{\sigma_g(x)}
 &=\rho(h_{\sigma_g(x)})^{-1}
   \left(\rho_{\mathrm{glob}}(g)J\psi\right)_{\sigma_g(x)} \\
 &=\rho(h_{\sigma_g(x)})^{-1}
   \rho(\kappa(g,x))\rho(h_x)\psi_x \\
 &=\left(\rho'_{\mathrm{glob}}(g)\psi\right)_{\sigma_g(x)}.
\end{align*}
Hence, for every $g\in T$,
\begin{equation}
  \rho'_{\mathrm{glob}}(g)
  =J^{-1}\rho_{\mathrm{glob}}(g)J.
\end{equation}
\end{proof}

\section{Adapted bases and reduced endpoint blocks}
\label{app:adapted-bases}
For the finite site basis used in the numerical examples, an orthonormal
basis for the range of an orthogonal projector is obtained from its
eigenvectors with eigenvalue one. Concatenating such columns for the
global projectors $\{\mathcal P_\gamma\}$ gives a unitary matrix $S$,
with each isotypic component occupying a consecutive block of columns.
The Hamiltonian in this adapted basis is $S^\dagger H S$, as shown in
Figure~\ref{fig:square-hamiltonian}.

For a local endpoint, let $L_{x,\lambda}$ have orthonormal columns spanning
$\operatorname{Ran}\overline{\mathcal Q}_{x,\lambda}$. Then
\[
  L_{x,\lambda}^\dagger L_{x,\lambda}=I,
  \qquad
  L_{x,\lambda}L_{x,\lambda}^\dagger
  =\overline{\mathcal Q}_{x,\lambda}.
\]
The reduced block from $(x_j,\mu)$ to $(x_i,\lambda)$ is
\[
  L_{x_i,\lambda}^\dagger H L_{x_j,\mu}.
\]
Its Hilbert--Schmidt norm equals
$\|\overline{\mathcal Q}_{x_i,\lambda}
H\overline{\mathcal Q}_{x_j,\mu}\|_{\mathrm{HS}}
$, because multiplication by the endpoint isometries
preserves the nonzero singular values. This gives the channel norms in
Figure~\ref{fig:coupling-dynamics}a without retaining zero rows and columns
from the complementary sectors.

\section{Reynolds averaging as an orthogonal projection}
\label{app:reynolds-projection}
The averaging map of Section~\ref{sec:operator-symmetrization} is an
orthogonal projection with respect to the Hilbert--Schmidt inner product
\[
  \langle A,B\rangle_{\mathrm{HS}}
  :=\operatorname{Tr}(A^\dagger B).
\]
For finite-dimensional $W$, this inner product is defined on
$\operatorname{End}(W)$. More generally, it is defined on the Hilbert
space $\mathcal B_2(W)$ of Hilbert--Schmidt operators. In that setting,
the orthogonal projection has image
$\mathcal B_2(W)\cap\operatorname{End}_T(W)$.

Write $U_g=\rho_{\mathrm{glob}}(g)$. Conjugation by $U_g$ is unitary
on the Hilbert--Schmidt operator space, since
\[
  \langle U_gAU_g^\dagger,U_gBU_g^\dagger\rangle_{\mathrm{HS}}
  =\operatorname{Tr}(U_gA^\dagger BU_g^\dagger)
  =\operatorname{Tr}(A^\dagger B).
\]
For finite $T$, the Reynolds average is
\[
  \mathcal R_T(B)=\frac{1}{|T|}\sum_{h\in T}U_hBU_h^\dagger.
\]
Reindexing by $h\mapsto gh$ gives
$U_g\mathcal R_T(B)U_g^\dagger=\mathcal R_T(B)$.
Thus $\mathcal R_T(B)$ commutes with the representation. Conversely,
if $A$ commutes with every $U_h$, then $\mathcal R_T(A)=A$.
These identities establish the image and idempotence of the average.

To establish orthogonality, let $A$ be an invariant Hilbert--Schmidt
operator and let $B$ be any Hilbert--Schmidt operator. Then
\begin{align*}
  \langle A,\mathcal R_T(B)\rangle_{\mathrm{HS}}
  &=\frac{1}{|T|}\sum_{h\in T}
    \operatorname{Tr}(A^\dagger U_hBU_h^\dagger)\\
  &=\frac{1}{|T|}\sum_{h\in T}
    \operatorname{Tr}(U_h^\dagger A^\dagger U_hB)\\
  &=\langle A,B\rangle_{\mathrm{HS}},
\end{align*}
where the final equality uses the invariance of $A^\dagger$.
Consequently,
\[
  \langle A,B-\mathcal R_T(B)\rangle_{\mathrm{HS}}=0.
\]
The residual is orthogonal to every invariant operator, proving the
orthogonal-projection assertion. For compact $T$, the same argument
uses normalized Haar integration in place of the finite average.

\section{Energy scaling of the illustrated dynamics}
\label{app:energy-scaling}
For the self-adjoint Hamiltonian $H$ on the $N$-dimensional aggregate
space $W$, we define
\begin{equation}
\label{eq:spectral-energy-scale}
  E_*:=\sqrt{\frac{1}{N}\operatorname{Tr}\!\left[(H-\bar E I)^2\right]},
  \qquad
  \bar E:=\frac{\operatorname{Tr}H}{N}.
\end{equation}
This is the standard deviation of the eigenvalues of $H$, counting
multiplicity with equal weight. It provides a basis-independent energy
scale for the chosen Hamiltonian. Subtracting $\bar E I$ removes
the arbitrary energy origin, which affects the evolution only through a
common phase.

For $E_*>0$, introduce the dimensionless Hamiltonian and time
\[
  \widehat H:=\frac{H-\bar E I}{E_*},
  \qquad \tau:=\frac{E_*t}{\hbar}.
\]
Then
\[
  e^{-iHt/\hbar}
  =e^{-i\bar E t/\hbar}e^{-i\widehat H\tau}.
\]
The local populations in Figure~\ref{fig:coupling-dynamics}b are insensitive
to the common phase.
Plotting them against $\tau$ expresses the dynamics in units of
$\hbar/E_*$ and removes the arbitrary overall energy magnitude of the
randomly constructed Hamiltonian.

\end{document}